\documentclass[aps,prb,twocolumn,footinbib,showpacs,superscriptaddress,preprintnumbers,amsmath,amssymb]{revtex4-1}

\usepackage{graphicx}
\usepackage{subfigure}
\usepackage{dcolumn}
\usepackage{bm}
\usepackage{verbatim} 

\begin{document}

\title{Full-counting statistics of charge and spin transport in the transient regime: A nonequilibrium Green's function approach}

\author{Gao-Min Tang}
\author{Jian Wang}
 \email{jianwang@hku.hk}
\affiliation{Department of Physics and the Center of Theoretical and Computational Physics, The University of Hong Kong, Pokfulam Road, Hong Kong, China}

\date{\today}

\begin{abstract}
We report the investigation of full-counting statistics (FCS) of transferred charge and spin in the transient regime where the connection between central scattering region (quantum dot) and leads are turned on at $t=0$. A general theoretical formulation for the generating function (GF) is presented using a nonequilibrium Green's function approach for the quantum dot system. In particular, we give a detailed derivation on how to use the method of path integral together with nonequilibrium Green's function technique to obtain the GF of FCS in electron transport systems based on the two-time quantum measurement scheme. The correct long-time limit of the formalism, the Levitov-Lesovik's formula, is obtained. This formalism can be generalized to account for spin transport for the system with noncollinear spin as well as spin-orbit interaction. As an example, we have calculated the GF of spin-polarized transferred charge, transferred spin, as well as the spin transferred torque for a magnetic tunneling junction in the transient regime. The GF is compactly expressed by a functional determinant represented by Green's function and self-energy in the time domain. With this formalism, FCS in spintronics in the transient regime can be studied. We also extend this formalism to the quantum point contact system. For numerical results, we calculate the GF and various cumulants of a double quantum dot system connected by two leads in transient regime. The signature of universal oscillation of FCS is identified. On top of the global oscillation, local oscillations are found in various cumulants as a result of the Rabi oscillation. Finally, the influence of the temperature is also examined.
\end{abstract}

\maketitle

\section{Introduction}
A stochastic process can be characterized by the distribution function. In many cases, the distribution function of a physical quantity is Gaussian and hence only two variables are enough to describe the distribution: its average and second cumulant. Due to the particle nature and quantum effect, electron noise spectrum is an intrinsic property that manifests in mesoscopic systems.\cite{Blanter} It was predicted theoretically that distribution of electron current is binomial, suggesting that all cumulants of current have to be included in order to fully characterize the electronic quantum transport process.\cite{Levitov1,Levitov2} The full-counting statistics (FCS) is an elegant way to study the current correlations in mesoscopic systems and yield not only the noise but all higher order cumulants.\cite{Levitov3} It calculates the probability distribution function of the number of electrons transferred through a particular terminal during given period of time that contains fundamental information about the current fluctuation in the system.\cite{Kampen}
The current and its fluctuations in mesoscopic systems have been studied extensively and are very important to characterize the physical mechanisms and correlations of a quantum transport systems.\cite{Blanter} For instance, the effective charge of quasi-particle can be determined from shot noise measurement in fractional quantum Hall effect.\cite{sam}
The cross current correlation can reveal statistical information such as whether the quasi-particle is fermionic or bosonic.
The study of correlation of entangled electron can be valuable in quantum information processing.\cite{loss}
A deep relationship has been found between entanglement and noise in terms of FCS providing new framework for quantum entanglement.\cite{klich} Furthermore, the equivalence between fidelity of quantum systems and generating function for FCS provides a link between fields of quantum transport and quantum information.\cite{lesovik} In addition, the measurement of cumulants to very high orders has been carried out experimentally for electronic transport in quantum point contact systems.\cite{exp1,exp2,exp3}So far, extensive investigation has been carried out on the FCS of charge transport, less attention has been paid to FCS of spin transport. It is the purpose of this paper to address this problem.

The key of FCS is to obtain the generating function (GF) from which the probability distribution $P(n,t)$ and all cumulants are calculated.\cite{Kampen} The GF can be calculated by various ways. Using a gedanken experiment scheme of a "charge counter" in the form of spin precession, Levitov and Lesovik,\cite{Levitov1,Levitov2,Levitov3} gave an analytical expression for the GF in the long-time limit which can be generalized to a general quantum mechanical variable.\cite{Nazarov} The GF has been obtained using the first quantization method\cite{wavepacket} which can be used to study FCS of dc and ac transport.\cite{albert2,flindt3} Using the nonequilibrium Green's function (NEGF)\cite{19,20} and path integral method (PI) in the two-time quantum measurement scheme\cite{RMP,twotime1,twotime2, twotime3,Schonhammer}, the GF has been calculated to study FCS of phonon transport\cite{JS1,JS2,JS3,JS4} and electric transport.\cite{gm}

In this paper, we generalize the existing formalism of FCS of charge transport in the two-measurement scheme to spin transport in the transient regime. In particular, we obtain GFs for spin polarized charge current, spin current, and spin transfer torque in the transient regime for a magnetic tunneling junction where the spin index is not a good quantum number. We have also extended this NEGF-PI method to quantum point contact systems for charge transport. As an application for this formalism, numerical results are given for FCS of charge transport in transient regime for a double quantum dot system.
	
The paper is organized as follows. In Sec. II, we give a basic definition of quantities needed in studying FCS, and in Sec. III, which is the central part of this paper, we present details on how to use the method of path integral together with NEGF to calculate the GF of FCS for lead-QD-lead system based on the two-time quantum measurement scheme. This formalism is designed for transient dynamics. The generalization of this formalism to spintronics in transient regime is provided in Sec. IV where we use the magnetic tunnel junction (MTJ) as an example. The GF for spin polarized charge transport, spin transport and spin transferred torque for MTJ are calculated. In Sec. V, we generalize the formalism to the quantum point contact system. Sec. VI is devoted to some numerical results where we apply the formalism to calculate various cumulants of transferred charge for a double quantum dot system. Finally concluding remarks are made in Sec. VII.

\section{Statistics}
The most important quantity in FCS is the GF, from which various quantities of interest can be obtained. In general GF is denoted as $Z(\lambda,t)$ where $\lambda$ is the counting field. The GF is defined as the Fourier transform of the probability distribution $P(\Delta n,t)$ of the number of transferred electrons $\Delta n=n_t-n_0$ which can be calculated from two-time quantum measurement scheme between time $t_0=0$ and $t$,\cite{Schonhammer}
\begin{equation} \label{eq1}
Z(\lambda,t)\equiv\left< e^{i\lambda\Delta n}\right>=\sum_{\Delta n}P(\Delta n,t)e^{i\lambda \Delta n} ,
\end{equation}
where $\Delta n$ can be either positive or negative. Various moments of transferred charge $\left<(\Delta n)^j\right>$ can be obtained by expanding $Z(\lambda,t)$ in terms of $\lambda$, we have
\begin{equation}  \label{eq2}
Z(\lambda,t)=\sum_{j=0}^{\infty}\frac{(i\lambda)^j}{j!}\left<(\Delta n)^j\right> .
\end{equation}
The $j$th cumulant $\langle\langle(\Delta n)^j\rangle\rangle$ can be calculated by taking the $j$th derivative of the cumulant generating function (CGF) which is the logarithm of GF with respect to $\lambda$ at $\lambda=0$:
\begin{equation} \label{eq3}
\langle\langle(\Delta n)^j\rangle\rangle=\frac{\partial^j \ln Z(\lambda,t)}{\partial(i\lambda)^j}\bigg|_{\lambda=0} .
\end{equation}

It is well known that cumulants can be expressed by moments. For instance, the first cumulant (mean value) is defined as $\langle\langle\Delta n\rangle\rangle=\left<\Delta n\right>$, the second cumulant (variance) is given by $\langle\langle(\Delta n)^2\rangle\rangle=\left<(\Delta n)^2\right>-\left<\Delta n\right>^2$, and the third cumulant (skewness) is $\langle\langle(\Delta n)^3\rangle\rangle=\left<(\Delta n-\left<\Delta n\right>)^3\right>$.

With the GF, the distribution function for the number of the electrons $P(\Delta n,t)$ can be found through
\begin{equation} \label{eq4}
P(\Delta n,t)=\int_0^{2\pi}\frac{d\lambda}{2\pi} Z(\lambda,t) e^{-i\lambda\Delta n} .
\end{equation}
In particular, the idle time probability, the probability of no electrons measured at time $t$ is
\begin{equation}  \label{eq5}
\Pi(t)=P(0,t)=\int_0^{2\pi}\frac{d\lambda}{2\pi} Z(\lambda,t)  ,
\end{equation}
from which we can calculate the waiting times distribution for the electronic transport system in the transient regime.\cite{gm}

Now let us turn to the discussion of waiting time distribution (WTD). In the dc steady state transport, the WTD can be calculated through\cite{albert2} $\mathcal{W}_2(t) =\langle t \rangle \frac{d^2 \Pi(t)}{dt^2} $ where $\langle t \rangle$ is the averaged time and WTD depends only on $t$ because of the time-translational invariance in the dc case (steady state). In the presence of ac bias, averaging over a time period is needed so that WTD depends only on $t$ as well.\cite{flindt3} However, in the transient transport regime, time translational invariance does not exist and there is also no time periodicity like the ac case. As discussed in details in Ref.\onlinecite{gm} that in the transient regime, we ask how long we wait for the detection of the first transferred electron if we set $t_0=0$ as the starting point. We will use $\mathcal{W}_1$ to denote the WTD in the transient regime\cite{Kampen,gm}
\begin{equation} \label{eq6}
\mathcal{W}_1(t)=-\frac{d}{dt}\Pi(t) .
\end{equation}

\section{Model and Generating Function}

\begin{figure}
  \includegraphics[width=3.0in]{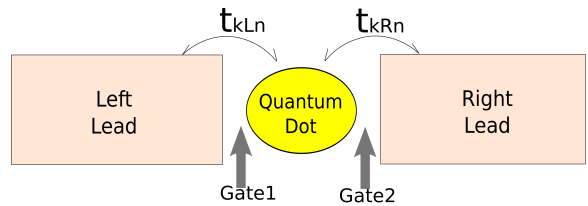}\\
  \caption{Schematic diagram of the model: a central quantum dot coupled to the left and right lead. The two gates between the leads and the central quantum dot can be used to control the coupling between the leads and the quantum dot. In the theoretical derivation, we count the number of transferred electrons in the left lead. }
  \label{fig1}
\end{figure}

\subsection*{a. Two-time quantum measurement}

We consider the system of a quantum dot denoted by $S$ connected by the left and right leads. The full Hamiltonian of the whole system can be written as
\begin{equation}  \label{eq7}
H=H_0+H_T=H_{L}+H_{R}+H_{S}+H_T
\end{equation}
where $H_0$ consists of the Hamiltonian of the isolated leads and the isolated central quantum dot,
\begin{equation}  \label{eq8}
H_S=\sum_{x\in n} \epsilon_{x} c_{x}^{\dag}c_{x}, \quad H_\alpha=\sum_{x\in k\alpha} \epsilon_{x} c_{x}^{\dag}c_{x},
\end{equation}
where we used the index $k\alpha$ to label the states of the lead $\alpha=L,R$ and the index $n$ for that of the quantum dot $S$. Here $\epsilon_{k\alpha}=\epsilon_{k\alpha}^{(0)}+q\Delta_{\alpha}$, where $\epsilon_{k\alpha}^{(0)}$ is the energy levels in the lead $\alpha$ and $\Delta_{\alpha}$ is the external voltage, $\epsilon_n$ is the energy levels of the quantum dot and $H_T$ is the Hamiltonian describing the coupling between the two leads and the quantum dot with the coupling constant $t_{k\alpha n}$,
\begin{equation}  \label{eq9}
H_T=H_{LS}+H_{RS}=\sum_{k\alpha n}[t_{k\alpha n}c_{k\alpha}^{\dag}c_{n}+t_{nk\alpha}c_{n}^{\dag}c_{k\alpha} ]
\end{equation}
where $t_{nk\alpha}=t_{k\alpha n}^*$. The coupling between the two leads and the quantum dot can be controlled by the two gates between the leads and the central quantum dots as shown schematically in Fig.1. 

To investigate full-counting statistics, we count the number of transferred electrons in the left lead, and the electrons flowing from the left lead to the quantum dot is defined as positive direction of the current. The current operator is given by ($q=1$)
\begin{equation} \label{eq10}
\hat{I}_L(t)=-\frac{d N^{(h)}_L(t)}{d t},
\end{equation}
where $N_L^{(h)}(t)=\sum_{k}c^\dag_{kL}(t)c_{kL}(t)$ is the electron number operator in the L-lead, and the superscript '($h$)' denotes the Heisenberg picture. $N_L^{(h)}(t)$ is related to the number operator in the Schrodinger picture $N_L(0)$ by,
\begin{equation}    \label{eq11}
N^{(h)}_L(t)=U(0,t)N_L(0)U(t,0)
\end{equation}
where the evolution operator is
\begin{equation} \label{eq12}
U(t,t')=\mathcal{T}\exp\left\{-\frac{i}{\hbar}\int_{t'}^tH(t_1)dt_1\right\}, \quad (t>t'),
\end{equation}
and $\mathcal{T}$ is the time-ordering operator. The anti-time ordering operator $\mathcal{\widetilde{T}}$ should be used if $t<t'$ and $U^\dag(t,t')=U(t',t)$.
From the Heisenberg's equation of motion $\mathrm{d}A/\mathrm{d}t=-\frac{i}{\hbar}[A,H]$, we find,
\begin{equation} \label{eq13}
\hat{I}_L(t)=\frac{i}{\hbar}[N^{(h)}_L(t),H(t)]=\frac{i}{\hbar}\sum_{kn}t_{kLn}c^\dag_{kL}(t)c_n(t)+H.c..
\end{equation}

Now we discuss the two-time quantum measurement by counting the number of electrons in the L-lead. In the two-time measurement scheme we measure the physical quantity such as number operator $N_L$ at two different times, e.g., first at time $0$ and then at time $t$. After each measurement, the system is projected onto one of the eigenstates of the operator $N_L$ with the corresponding eigenvalue.  We define the projection operator at time $0$ and $t$ as $P_0$ and $P_t$, respectively. Let us start from an initial state $|\Psi_0\rangle$ and assume that $|n_0\rangle$ forms a complete set of eigenstates of number operator at time $t=0$, we have
\begin{equation} \label{eq14}
N_L(0)|n_0\rangle=n_0|n_0\rangle, \ P_0=|n_0\rangle\langle n_0|.
\end{equation}
Obviously we have ${P_{0}^2}=P_{0}$ and $\sum_{n_0}P_{0}=1$ and similar relations hold for $P_t$.

After the first measurement at time 0, the wave function becomes $P_0|\Psi_0\rangle$ with a probability of finding this state equal to $\langle\Psi_0|P_0^2|\Psi_0\rangle$. After a time interval t, this state evolves to a new state $U(t,0)P_0|\Psi_0\rangle$ with an eigenvalue $n_t$. After the second measurement at time t, the wave function becomes $|\Psi_t\rangle=P_tU(t,0)P_0|\Psi_0\rangle$, where $P_t = |n_t\rangle\langle n_t|$.

Assuming that the initial state is a mixed state with the density operator,
\begin{equation} \label{eq15}
\rho(0)=\sum_{k}\omega_k|\psi_0^k\rangle\langle\psi_0^k|, \; \sum_{k}\omega_k=1  ,
\end{equation}
we find the joint probability to have measured $n_0$ electrons at time $0$ and $n_t$ electrons at time $t$,
\begin{align}   \label{eq16}
P(n_t,n_0)&=\sum_k \omega_k\langle\psi_0^k|P_0U(0,t)P_t^2U(t,0)P_0|\psi_0^k\rangle   \notag  \\
&=\mathrm{Tr}[P_0\rho(0)P_0U(0,t)P_tU(t,0)]
\end{align}
Keep in mind that we should add a normalization constant to the joint probability and the GF, Eq.~(\ref{eq23}). We will normalize the GF when we come to the final result and use the fact that $Z(\lambda=0)=1$.  The probability distribution for the number of electrons $\Delta n=n_t-n_0$ measured between two measurements is given by
\begin{equation}  \label{eq17}
P(\Delta n)=\sum_{n_t,n_0}\delta[\Delta n-(n_0-n_t)]P(n_t,n_0),
\end{equation}
where $\delta(n)$ is the Kronecker $\delta$ symbol. Using Eq.~(\ref{eq14}), we have $n_0P_0=N_L(0)P_0$ and $n_tP_t=N_L(0)P_t$. The GF associated with the probability $P(\Delta n)$ is\cite{RMP,foot1}
\begin{align}   \label{eq18}
&Z(\lambda,t) \equiv \sum_{\Delta n}P(\Delta n)e^{i\lambda \Delta n}=\sum_{n_t,n_0}e^{i\lambda(n_0-n_t)}P(n_t,n_0)   \notag  \\
&=\sum_{n_t,n_0}\mathrm{Tr}[e^{i\lambda N_L(0)}P_0\rho(0)P_0U(0,t)e^{-i\lambda N_L(t)}P_tU(t,0)] \notag \\
&=\left\langle e^{i\lambda N_L(0)}e^{-i\lambda N_L^{(h)}(t)}\right\rangle^\prime \notag \\
&=\left\langle e^{i\lambda N_L(0)/2}e^{-i\lambda N_L^{(h)}(t)}e^{i\lambda N_L(0)/2}\right\rangle^\prime
\end{align}
where $P_t$ disappears after the summation over $n_t$ and the prime indicates that the average is with respect to
\begin{equation}  \label{eq19}
\rho^{\prime}(0)=\sum_{n_0}P_0\rho(0)P_0.
\end{equation}

To remove the projection operator $P_0$, we represent it using the Kronecker delta function
\begin{align}  \label{eq20}
P_0&=|n_0\rangle\langle n_0| =\sum_n \int_0^{2\pi}\frac{d\xi}{2\pi}e^{-i\xi(n_0-n)}|n\rangle\langle n| \notag \\
&=\int_0^{2\pi}\frac{d\xi}{2\pi}e^{-i\xi(n_0-N_L(0))}  ,
\end{align}
then we can easily express $\rho'(0)$ in an integral form,
\begin{equation}  \label{eq21}
\rho'(0) =\int_0^{2\pi}\frac{d\xi}{2\pi} e^{i\xi N_L(0)}\rho(0)e^{-i\xi N_L(0)} .
\end{equation}


Using Eq.~(\ref{eq18}) and Eq.~(\ref{eq21}), we express the GF as follows,
\begin{align} \label{eq22}
Z(\lambda,t) =\int_0^{2\pi}\frac{d\xi}{2\pi}Z(\lambda,\xi,t) ,
\end{align}
with
\begin{equation}  \label{eq23}
Z(\lambda,\xi,t) =\mathrm{Tr}\left\lbrace\rho(0)U_{\lambda /2-\xi}(0,t)U_{-\lambda /2-\xi}(t,0)\right\rbrace,
\end{equation}
where $U_\gamma$ with $\gamma=\lambda /2-\xi$ or $-\lambda /2-\xi$ is the modified evolution operator ($t\geqslant t'$),
\begin{align} \label{eq24}
U_\gamma(t,t')
&=e^{i\gamma N_L(0)}U(t,t')e^{-i\gamma N_L(0)}  \notag \\
&=\mathcal{T}\exp\left\{-\frac{i}{\hbar}\int_{t'}^tH_\gamma(t_1)dt_1\right\}
\end{align}
with
\begin{equation} \label{eq25}
H_\gamma(t)=e^{i\gamma N_L(0)}H(t)e^{-i\gamma N_L(0)}.
\end{equation}
As mentioned before, the anti-time-ordering operator should be used here if $t<t'$.

\begin{figure}
  \includegraphics[width=3.50in]{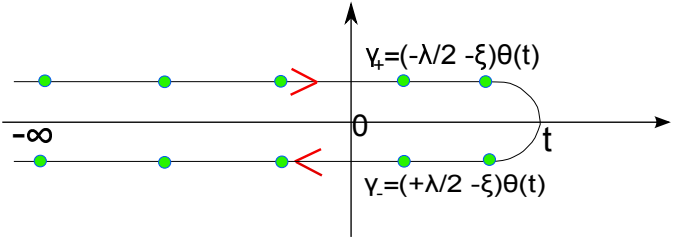}\\
  \caption{Complex contour defined from time $-\infty$ to time $t$ and then back to time $-\infty$ in Keldysh space. When we consider the case of the transient regime, in which the subsystems are connected at time $t_0=0$, the complex contour should be from time $t_0=0$ to time $t$ and then back to time $t_0=0$ in Keldysh space. }
  \label{fig2}
\end{figure}

Since $U_{-\lambda /2-\xi}(t,0)$ is from $0$ to $t$ and $U_{\lambda /2-\xi}(0,t)$ is from $t$ to $0$, we can use the Keldysh contour as shown in Fig.~(\ref{fig2}) to combine $U_{\lambda /2-\xi}(0,t)U_{-\lambda /2-\xi}(t,0)$, where for the upper branch of the Keldysh contour
\begin{equation}  \label{eq26}
\gamma_+(t)=(-\lambda/2-\xi) \theta(t),
\end{equation}
and for the lower branch
\begin{equation}  \label{eq27}
\gamma_-(t)=(\lambda/2-\xi) \theta(t),
\end{equation}
and $\theta(t)$ is the step function due to the fact that the first measurement starts at $t=0$. Note that for a time $t$ in the upper branch and a time $t'$ in the lower branch, we always have
$t<t'$. In terms of Keldysh contour, we can express $Z(\lambda,\xi,t)$ as:
\begin{equation} \label{eq28}
Z(\lambda,\xi,t)=\mathrm{Tr}\left\{\rho(0)\mathcal{T}_C\exp\left[-\frac{i}{\hbar}\int_C H_\gamma(t_1)dt_1 \right]\right\},
\end{equation}
where $\mathcal{T}_C$ is the contour-ordering operator on Keldysh contour which has upper and lower branches discussed above.
Noticing the fact that $N_L(0)$ commutes with every term except the coupling term $H_{LS}$ in Eq.(\ref{eq9}) and from the Baker-Hausdorff lemma
\begin{equation}\label{baker}
e^X Y e^{-X} =Y+[X,Y]+\frac{1}{2!}[X,[X,Y]]+\cdots ,
\end{equation}
we have $e^{i\gamma N_L(0)}c_{kL}e^{-i\gamma N_L(0)}=e^{-i\gamma N_L(0)}c_{kL}$, we obtain
\begin{align}  \label{eq29}
H_\gamma(t)=&\sum_{x\in k\alpha ,n}\epsilon_x c^\dag_x c_x+\sum_{kRn}t_{kRn}c^\dag_{kR} c_n +\sum_{kRn}t_{nkR}c^\dag_n c_{kR}  \notag  \\
&+e^{i\gamma}\sum_{kLn} t_{kLn}c^{\dag}_{kL} c_n+ e^{-i\gamma}\sum_{kLn} t_{nkL}c^{\dag}_n c_{kL} .
\end{align}
Note that in the modified Hamiltonian the counting field $\gamma$ only enters the coupling term between the central quantum dot and the L-lead where we count the number of electrons.

Consider a system where the interaction between the quantum dot and the two leads is adiabatically switched on from $t=-\infty$ to $t=0$, the nonequilibrium state $\rho(0)$ can be obtained by evolving the system from the initially decoupled state $\rho(-\infty)=\rho_L\otimes\rho_S\otimes\rho_R$ at $t=-\infty$. This process can be described by
\begin{equation}  \label{eq30}
\rho(0)=U(0,-\infty)\rho(-\infty)U(-\infty,0) ,
\end{equation}
We can rewrite Eq.~(\ref{eq23}) as
\begin{align} \label{eq31}
Z(\lambda,t)&=\int_0^{2\pi}\frac{d\xi}{2\pi}\mathrm{Tr}\left\lbrace\rho(-\infty)U_{\gamma_-}(-\infty,t)U_{\gamma_+}(t,-\infty)\right\rbrace   \notag  \\
&=\int_0^{2\pi}\frac{d\xi}{2\pi}Z(\lambda,\xi,t),
\end{align}
Similarly, in terms of Keldysh contour, we can express $Z(\lambda,\xi,t)$ as:
\begin{equation} \label{eq32}
Z(\lambda,\xi,t)=\mathrm{Tr}\left\lbrace\rho(-\infty)\mathcal{T}_K\exp\left\lbrace-i\int_K H_\gamma(t_1)dt_1 \right\rbrace\right\rbrace,
\end{equation}
where we have used 'K' to denote the contour, from $t_0=-\infty$ to $t$ and then back to $t_0=-\infty$, for this adiabatic process. In contrast, $t_0$ is $0$ in the previous contour 'C'. In general, we can discuss the following two initial conditions.\cite{JS2} \\
(1) Measurement Regime. The system starts at $t=-\infty$ with the three different regions (L,R,S) disconnected. The coupling between them and the dc bias voltage are switched on adiabatically after $t=-\infty$ and the system evolves to steady state up to time $t=0$. This is the dc transport regime and the current is independent of time in the steady state.
In this case we introduce projector $P_0$ to make the first measurement. This measurement is mathematically done by simply introducing a parameter $\xi$ in Eq.~(\ref{eq20}).   \\
(2) Transient regime. In this regime, the coupling between the leads and the quantum dot is switched on at $t_0=0^+$. As shown in Fig.1, two gate voltages are applied to control the coupling between leads and central scattering region. By changing these gate voltages, we can turn on and off the coupling at will. The density matrix at $t=0$, $\rho(0)$, is the product of initial states of decoupled subsystems $\rho(-\infty)=\rho_L\otimes\rho_S\otimes\rho_R$. We will see later that the above parameter $\xi$ will not appear under this regime. Obviously, the contour 'C' should be used in the transient regime. 

There is a fundamental question in FCS: how to probe the state of the system in a noninvasive way. As we will see below that after the first measurement, the quantum state of the system is altered. This means that the subsequence measurement will give a different result from what we should get if the system were not perturbed due to the first measurement. This has been noted in the early work of Levitov and Lesovik where a measuring device is attached to the right lead of the system so that the current can be measured from the rotating angle of the spin. As pointed out by Levitov and Lesovik, their measurement is non-invasive only in the sense that the reflection amplitude is unchanged and the transmission amplitude changes by a phase.\cite{Levitov3} However, there are important consequences due to this phase change. For instance, although the FCS at long times is correctly represented by the generating function derived from this approach, the second and higher order cumulants of current at short times will be different from the true value. This is a known problem in the FCS community.

The transient regime is different. This is because in the transient regime, the density matrix of the system at $t=0$ is a direct product of the three regions $\rho(0) = \rho_L \otimes\rho_D\otimes\rho_R$, the system will not be perturbed if the first measurement is performed at $t=0$. As a result the first measurement at $t=0$ is not necessary and hence only one measurement is needed. Therefore the quantum system will not be altered using the two-measurement scheme in the transient regime. 


\subsection*{b. Keldysh formalism}
Now we introduce the Keldysh formalism\cite{20,Ka1,Ka2} to derive GF. For this purpose it is convenient to use the Grassmann algebra whose basic knowledge is presented in Appendix A. We divide the Keldysh contour from $t=-\infty$ to $t$ and then back to $-\infty$ in Eq.~(\ref{eq31}) into $2N$ equal time intervals $\delta t$, such that $t_1=t_{2N}=-\infty$ and  $t_N=t_{N+1}=t$. We will use the relation of over-completeness of the Fermion coherent state Eq.~(\ref{eqA15}) and insert it at each time slice $i=1,2,...,2N$ along the contour.\cite{Ka1,Ka2} It is important to note that the Grassmann fields $\overline{\phi}$ and $\phi$ are completely independent fields.
Introducing the abbreviation for evolution operator over $\delta t$, $U(\delta t_j)\equiv U(t_0+ j \delta t,t_0+ (j-1)\delta t)$ and using Eq.~(\ref{eqA13}),
we find ($\hbar=1$):
\begin{align}  \label{eq33}
&\langle\phi_{j+1}|U(\delta t_j)|\phi_{j}\rangle=\exp\left\{-i\delta t_j\left[\sum_{x\in k\alpha, n}\epsilon_x \overline{\phi}_{(j+1)x}\phi_{jx}
\right. \right. \notag  \\
&+\sum_{kL,n} \left( e^{i\gamma_j}t_{kLn}\overline {\phi}_{(j+1)kL}\phi_{jn}+e^{-i\gamma_j} t_{nkL} \overline{\phi}_{(j+1)n}\phi_{jkL} \right)      \notag  \\
&\left. \left. +\sum_{kR,n} \left( t_{kRn}\overline{\phi}_{(j+1)kR}\phi_{jn}+t_{nkR}
\overline{\phi}_{(j+1)n}\phi_{jkR} \right) \right] \right\} \langle\phi_{j+1}|\phi_{j}\rangle
\end{align}
where the $\delta t_j=+\delta t$ indicates the forward-time branch and $\delta t_j=-\delta t$ is for the backward-time branch and we use the index $k\alpha$ to label the states of the lead $\alpha$ and the index $n$ the quantum dot. Remember that $\gamma_j=\gamma_+$ if $j=1,2,\cdots ,N$ and  $\gamma_j=\gamma_-$ if $j=N,N+1,\cdots ,2N$. From Eq.~(\ref{eqA14}) one finds $\langle \phi_1| \hat{\rho}(-\infty)|-\phi_{2N}\rangle =\exp\{-\overline{\phi}_1\phi_{2N}\rho(-\infty) \} $. \cite{foot3}

Substituting Eq.~(\ref{eq33}) into Eq.~(\ref{eq32}) and using Eq.~(\ref{eqA16}) of the trace formula expressed in coherent states, we obtain the GF
\begin{equation} \label{eq34}
Z(\lambda,\xi,t)=\int\mathcal{D}[\overline{\phi}\phi]e^{iS[\overline{\phi}\phi]},
\end{equation}
with the action
\begin{align}  \label{eq35}
& S[\overline{\phi}\phi]=\sum_{j=1}^{2N-1}\left\{\sum_{x\in k\alpha, n}\overline{\phi}_{(j+1)x}\left[ i\frac{\phi_{(j+1)x}-\phi_{jx}}{\delta t_j}-\epsilon_x \phi_{jx}\right]  \right.  \notag \\
& - \sum_{kL,n} \left[ e^{i\gamma_j}\overline{\phi}_{(j+1)kL}t_{kLn}\phi_{jn}
+e^{-i\gamma_j} \overline{\phi}_{(j+1)n}t_{nkL}\phi_{jkL}\right] \notag \\
&\left. -\sum_{kR,n}\left[ \overline{\phi}_{(j+1)kR}t_{kRn} \phi_{jn}+ \overline{\phi}_{(j+1)n}t_{nkR}\phi_{jkR} \right]  \right\} \delta t_j   \notag \\
&+i\overline{\phi}_1(\phi_1 +\rho(-\infty)\phi_{2N}),
\end{align}
where the term $i\overline{\phi}_{(j+1)x}\phi_{(j+1)x}/\delta t_j$ in Eq.(\ref{eq35}) comes from the relation of over completeness of Fermion coherent states, Eq.~(\ref{eqA15}). The term $i\overline{\phi}_{(j+1)x}\phi_{(j)x}/\delta t_j$ in the above equation that contains two time indices is due to  $\langle\phi_{j+1}|\phi_{j}\rangle$ in Eq.(\ref{eq33}) after using Eq.~(\ref{eqA12}). To avoid integration along the closed time contour, we split the Grassmann field into upper and lower branches of the contour, respectively.\cite{Ka2} Here, we use $+$ and $-$ to differentiate the upper and lower branches. Setting $N\rightarrow\infty$ and $\delta t_j\rightarrow 0$, we can obtain the continuous expression for the action
\begin{align} \label{eq36}
&S[\overline{\phi}\phi]  \notag \\
=&\int_{-\infty}^t d\tau\sum_{x\in k\alpha, n}\left[\overline{\phi}_{x+}(i\partial_t-\epsilon_x)\phi_{x+}-\overline{\phi}_{x-}(i\partial_t-\epsilon_x)\phi_{x-}\right]    \notag \\
&-\sum_{k,n}\left[e^{i\gamma_+}\overline{\phi}_{kL+}t_{kLn}\phi_{n+}-
e^{i\gamma_-}\overline{\phi}_{kL-}t_{kLn}\phi_{n-}  \right. \notag  \\
&\left. \qquad \ +e^{-i\gamma_+}\overline{\phi}_{n+}t_{nkL}\phi_{kL+}-e^{-i\gamma_-}\overline{\phi}_{n-}t_{nkL}\phi_{kL-} \right] \notag  \\
&-\sum_{k,n}\left[\overline{\phi}_{kR+}t_{kRn}\phi_{n+}
-\overline{\phi}_{kR-}t_{kRn}\phi_{n-}    \right. \notag  \\
&\left. \qquad  \ +\overline{\phi}_{n+}t_{nkR}\phi_{kR+}-
\overline{\phi}_{n-}t_{nkR}\phi_{kR-} \right].
\end{align}
The last term $-\rho(-\infty)$ in Eq.~(\ref{eq35}) is responsible for the boundary condition at the $-\infty$ to connect the upper and lower branches and this will be easily seen in Eq.~(\ref{eq39}) later. \cite{Ka2}

Now, we want to express Eq.~(\ref{eq36}) in terms of Keldysh Green's function. To do that, we consider the free action of the quantum dot or the leads in the absence of coupling between them or external fields in Eq.~(\ref{eq35})
\begin{align} \label{eq37}
S_0=& \sum_{j=1}^{2N-1} \overline{\phi}_{(j+1)}\left[ i\frac{\phi_{(j+1)}-\phi_{j}}{\delta t_j}-\epsilon \phi_{j}\right]{\delta t_j}  \notag \\
& +i\bar{\phi}_1(\phi_1 +\rho(-\infty)\phi_{2N})
\equiv\sum_{j,j'}^{2N}\bar{\phi}_j g_{jj'}^{-1}\phi_{j'}  .
\end{align}
where $g_{jj'}^{-1}$ has double time indices. From the basic property of the Gaussian integral for Grassmann algebra we have
\begin{equation} \label{eq38}
\langle \phi_j\bar{\phi}_{j'}\rangle =\frac{\int \mathcal{D}[\bar{\phi}\phi] \phi_j\bar{\phi}_{j'} \exp(iS_0)}{\int \mathcal{D}[\bar{\phi}\phi] \exp(iS_0)}=ig_{jj'}
\end{equation}
From Eq.~(\ref{eq37}) we can write the matrix $ig_{jj'}^{-1}$ in the following form (when $N=3$)
\begin{equation} \label{eq39}
i g_{jj'}^{-1}=
\left(
  \begin{array}{ccc}
    -1  &      &    \\
    h_- & -1   &    \\
        & h_-  &  -1    \\
        \hline
      &    &   1    \\
     &   &    \\
        &   &
  \end{array}
  \vline
    \begin{array}{ccc}
      &  & -\rho \\
      &  &  \\
          &  & \\
          \hline
       -1 &  &  \\
     h_+ & -1 &  \\
        & h_+ & -1
  \end{array}
\right)
\end{equation}
where $h_{\pm}\equiv 1 \mp i\epsilon \delta_t$. As shown in Ref.\onlinecite{Ka1,Ka2}, we can get the discrete form Green's function of the free quantum dot or the lead by inverting the matrix in Eq.(\ref{eq39}). The continuous version of the Green's function can be obtained by taking the $N\rightarrow \infty$ limit while keeping $N\delta_t$ constant and also $(h_+ h_-)^N \rightarrow 1$.
Then the four correlation functions in the continuum limit are\cite{Ka2}
\begin{align}  \label{eq40}
& \langle \phi^+(t)\bar{\phi}^-(t') \rangle =i g^<(t,t')= -n_F\exp\{-i\epsilon (t-t')\} \notag  \\
& \langle \phi^-(t)\bar{\phi}^+(t') \rangle =i g^>(t,t')= (1-n_F)\exp\{-i\epsilon (t-t')\} \notag  \\
& \langle \phi^+(t)\bar{\phi}^+(t') \rangle =i g^t(t,t')= \theta(t-t')ig^>+\theta(t'-t)ig^< \notag  \\
& \langle \phi^-(t)\bar{\phi}^-(t') \rangle =i g^{\bar{t}}(t,t')= \theta(t'-t)ig^> +\theta(t-t')ig^<
\end{align}
where $n_F=\rho/(1+\rho)$ is the Fermi occupation number.

Now we perform the Keldysh rotation. Define the new fields as:
\begin{equation}  \label{eq41}
\phi_{a1}=\frac{1}{\sqrt{2}}(\phi_{a+}+\phi_{a-});\;
\phi_{a2}=\frac{1}{\sqrt{2}}(\phi_{a+}-\phi_{a-}),
\end{equation}
whereas "bar" fields transform differently:
\begin{equation}  \label{eq42}
\overline{\phi}_{a1}=\frac{1}{\sqrt{2}}(\overline{\phi}_{a+}-\overline{\phi}_{a-});\;
\overline{\phi}_{a2}=\frac{1}{\sqrt{2}}(\overline{\phi}_{a+}+\overline{\phi}_{a-}).
\end{equation}
The effect of this rotation is to transform the matrix form of contour-ordered function $A$ into an upper triangular matrix as follows:
\begin{equation}  \label{eq43}
\left(
  \begin{array}{cc}
    A^{t}(t,t') & A^{<}(t,t')   \\
    A^{>}(t,t') & A^{\bar{t}}(t,t')
  \end{array}
\right)
\longrightarrow
\left(
  \begin{array}{cc}
    A^r(t,t')   &  A^k(t,t')   \\
          0		 &  A^a(t,t')
  \end{array}
\right)
\end{equation}
with the following relation:
\begin{align}  \label{eq44}
&\left(
  \begin{array}{cc}
    A^r          &  A^k    \\
    0  &  A^a
  \end{array}
\right)
= Q \sigma_z
\left(
  \begin{array}{cc}
    A^{t}  &  A^{<}   \\
    A^{>}  &  A^{\bar{t}}
  \end{array}
\right) Q^T        \notag  \\
&=\frac{1}{2}
\left(
  \begin{array}{cc}
    A^t-A^{\bar{t}}-A^{<}+A^{>}  &  A^t+A^{\bar{t}}+A^{<}+A^{>}   \\
    A^t+A^{\bar{t}}-A^{<}-A^{>}  &  A^t-A^{\bar{t}}+A^{<}-A^{>}
  \end{array}
\right) ,
\end{align}
where $Q=\frac{1}{\sqrt{2}}
\left(
  \begin{array}{cc}
    1 &  -1   \\
    1 &  1
  \end{array}
\right)
$
and
$\sigma_z= \left(
  \begin{array}{cc}
    1 &  0   \\
    0 & -1
  \end{array}
\right)
$ are orthogonal matrices. Here, $A^r(t,t')$ and $A^a(t,t')$ are, respectively, the usual retarded and advanced Green's function. For Green's functions or self-energies without counting parameter or other parameters involved, we have
\begin{align}   \label{eq45}
A^t+A^{\bar{t}} &=A^{<}+A^{>} ,   \notag  \\
A^k &=2A^{<}+A^{r}-A^{a}    .
\end{align}
Introducing 
\begin{equation}  
 \begin{array}{cc}
  \overline{\psi}_x^T(\tau)=\left(\overline{\phi}_{x+}(\tau),\,\overline{\phi}_{x-}(\tau)  \right),  &
   \psi_x^T(\tau)=\left({\phi}_{x+}(\tau),\,{\phi}_{x-}(\tau)\right)
 \end{array} \nonumber
\end{equation}
and 
\begin{equation} 
 \begin{array}{cc}
  \overline{\phi}_x^T(\tau)=\left(\overline{\phi}_{x1}(\tau),\,\overline{\phi}_{x2}(\tau)  \right),  &
   \phi_x^T(\tau)=\left({\phi}_{x1}(\tau),\,{\phi}_{x2}(\tau)\right)
 \end{array} \nonumber
\end{equation}
we have from Eq.(\ref{eq41}) and (\ref{eq42})
\begin{eqnarray}
&&\overline{\psi}_x=Q \overline{\phi}_x \nonumber \\
&&\psi_x= \sigma_z Q \phi_x \label{eq44a}
\end{eqnarray}
The second and third terms of Eq.(\ref{eq36}) (denoted as $S_1$) can be written as
\begin{align} \label{eq36a}
&S_1[\overline{\phi}\phi]  \notag \\
=&-\int_{-\infty}^t d\tau \sum_{k,n}\left[ t_{kLn}\overline{\psi}_{kL}^T {\bf V} \psi_{n} + t_{nkL}\overline{\psi}_{n}^T {\bf V}^* \psi_{kL}\right. \notag  \\
&\left.+ t_{kRn}\overline{\psi}_{kR}^T \sigma_z \psi_{n} + t_{nkR}\overline{\psi}_{n}^T \sigma_z \psi_{kR}\right].
\end{align}
where ${\bf V} = (e^{i\gamma_+}-e^{i\gamma_-})/2+\sigma_z (e^{i\gamma_+}+e^{i\gamma_-})/2$. Substituting Eq.(\ref{eq44a}) into (\ref{eq36a}), we can rewrite the action of Eq.~(\ref{eq36}) after Keldysh rotation as follows:
\begin{align} \label{eq46}
&S[\overline{\phi}\phi]=S_{L}+S_{R}+S_{S}+S_{LS}+S_{RS}  \notag  \\
=&\int_{-\infty}^td\tau\int_{-\infty}^td\tau' \sum_{x,x'\in k\alpha n}\overline{\phi}_x^T(\tau) g_{xx'}^{-1}(\tau,\tau') \phi_{x'}(\tau')  \notag \\
-&\sum_{k,n}\left[t_{kLn}\overline{\phi}_{kL}^T(\tau)\Lambda(\gamma) \phi_{n}(\tau)
+t_{nkL}\overline{\phi}_n^T(\tau)\Lambda^*(\gamma) \phi_{kL}(\tau) \right]  \notag \\
-&\sum_{k,n}\left[t_{kRn}\overline{\phi}_{kR}^T(\tau) \phi_{n}(\tau)
+t_{nkR}\overline{\phi}_{n}^T(\tau)\phi_{kR}(\tau) \right]  ,
\end{align}
where we have introduced the abbreviated notation
\begin{align}  \label{eq48}
\Lambda(\gamma)&=Q^T {\bf V} \sigma_z Q \notag \\
&=\left(
  \begin{array}{cc}
    (e^{i\gamma_+}+e^{i\gamma_-})/2 & (e^{i\gamma_+}-e^{i\gamma_-})/2   \\
    (e^{i\gamma_+}-e^{i\gamma_-})/2 & (e^{i\gamma_+}+e^{i\gamma_-})/2
  \end{array}
\right)    \notag \\
&=
\begin{cases}
\exp(-i\xi)\exp\left(-\frac{i\lambda}{2}\sigma_x  \right) , & \tau \geq 0  \\
\mathrm{identity \ matrix},  & \tau <0
\end{cases}
\end{align}
with $\exp\left(-\frac{i\lambda}{2}\sigma_x \right)= \left(
  \begin{array}{cc}
    \cos\frac{\lambda}{2} & -i\sin\frac{\lambda}{2}   \\
    -i\sin\frac{\lambda}{2} & \cos\frac{\lambda}{2}
  \end{array}  \right) $. Here, the Green's function in Keldysh formalism is given by
\begin{equation}  \label{eq49}
g_{xx'}(t,t')=\left(
  \begin{array}{cc}
    g_{xx'}^{r}(t,t')  &  g_{xx'}^{k}(t,t')   \\
    0                        &  g_{xx'}^{a}(t,t')
  \end{array}
\right) ,
\end{equation}
where $(i\partial_t-\epsilon_x)g_{xx'}^{k}(t,t')=0$ and $(i\partial_t-\epsilon_x)g_{xx'}^{r,a}(t,t')=\delta(t-t')\delta_{x,x'}$. We point out that the coupling coefficients $t_{kLn},t_{kLn}^*,t_{kRn},t_{kRn}^*$ can also depend on $\tau$ in Eq.~(\ref{eq46}).

Now, we write Eq.~(\ref{eq46}) in a matrix form
\begin{equation} \label{eq50}
S[\overline{\phi}\phi]=\int_{-\infty}^td\tau \int_{-\infty}^td\tau'
\overline{\bf \Phi}^{T}(\tau) \mathcal{M}(\tau,\tau') {\bf \Phi}(\tau'),
\end{equation}
where we have used the notation $\overline{\bf \Phi}^{T}(\tau)=(\overline{\phi}_{kL}^{T}(\tau), \overline{\phi}_{n}^{T}(\tau),\overline{\phi}_{kR}^{T}(\tau)) $ and
\begin{equation} \label{eq51}
\mathcal{M}=
\left(
  \begin{array}{ccc}
    g_{kk' L}^{-1}(\tau,\tau') & -t_{kLn'}(\tau,\tau)\Lambda (\tau) & 0 \\
   -\Lambda^*(\tau)t_{nk'L}(\tau,\tau) & g_{nn'}^{-1}(\tau,\tau')
	& -t_{nk'R}(\tau,\tau) \\
   0  &  -t_{kRn'}(\tau,\tau) &  g_{kk' R}^{-1}(\tau,\tau')
  \end{array}
\right),
\end{equation}
where the matrix $\mathcal{M}(\tau,\tau')$ contains Keldysh time space, $k-$space, and orbital space. Note that $t_{k\alpha n}$ and $t_{nk\alpha}$ are diagonal matrices in Keldysh space which means that $t_{k\alpha n}^r=t_{k\alpha n}^a$ and $t_{k\alpha n}^<=0$. The upper bound for $\tau$ and $\tau'$ should be $t$, at which we take the second measurement.

Using functional integration of the Gaussian integral for independent Grassmann fields described by Eq.~(\ref{eqA11}) and taking into the normalization condition $Z(\lambda=0,t)=1$ and the fact $\Lambda(\lambda=0)=1$ into consideration, we can express the GF as follows,
\begin{equation}  \label{eq52}
Z(\lambda,t)=\frac{\det \mathcal{M}(\lambda)}{\det \mathcal{M}(\lambda=0)}  .
\end{equation}
Defining the diagonal matrix
\begin{equation}
\mathcal{P}=\left(  \begin{array}{ccc}
    g_{k'k L}(\tau,\tau')  & 0  &  0 \\
     0 &  I  & 0  \\
     0 &  0  &  g_{k'k R}(\tau,\tau')
  \end{array}  \right),
\end{equation}
we have
\begin{align}
\mathcal{PM}(\lambda)&=\left(  \begin{array}{ccc}
    1& -g_{k'k L}t_{kLn'}\Lambda &   0  \\
     -\Lambda^* t_{nk'L} &  g^{-1}_{n n'} & -t_{nk'R}  \\
      0 &  -g_{k'k R}t_{kRn'} &  1
  \end{array}  \right)  \notag \\
 &\equiv \left(  \begin{array}{ccc}
    A &  B  \\
    C &  D
  \end{array}  \right)
\end{align}
where
\begin{equation}
D \equiv \left(  \begin{array}{ccc}
      g^{-1}_{n n'} & -t_{nk'R}  \\
  -g_{k'k R}t_{kRn'} &  1
  \end{array}  \right)
\end{equation}
and $A=1$. Here the summation on repeated indices is implied. Using the identity
\begin{equation} \label{eqnew}
{\rm det}\left(  \begin{array}{ccc}
    A& B  \\
    C &  D
  \end{array}  \right) = {\rm det}(A) ~ {\rm det}(D-C A^{-1} B)
\end{equation}
we find
\begin{equation}
{\rm det}[\mathcal{PM}(\lambda)] = {\rm det}\left(  \begin{array}{ccc}
      g^{-1}_{n n'}-{\widetilde \Sigma}_L & -t_{nk'R}  \\
  -g_{k'k R}t_{kRn'} &  1
  \end{array}  \right)
\end{equation}
where
\begin{equation} \label{eq53}
\Sigma_L(\tau,\tau') =
\sum_{k,k'}t_{nk'L} g_{k'k L}(\tau,\tau')t_{k L n'}  ,
\end{equation}
and ${\widetilde\Sigma}_L(\tau,\tau')=\Lambda^*(\tau)\Sigma_L(\tau,\tau') \Lambda(\tau')$.
Using Eq.(\ref{eqnew}) again, we have ${\rm det}[\mathcal{PM}(\lambda)] = {\rm det}(g^{-1}_{nn'}-{\widetilde \Sigma}_L-{\Sigma}_R)$.
Finally from $Z(\lambda,\xi,t)=\det [\mathcal{P}\mathcal{M}(\lambda)]/\det [\mathcal{P}\mathcal{M}(\lambda=0)]$,
we obtain the normalized generating function in a compact form
\begin{equation}  \label{eq54}
Z(\lambda,\xi,t)=\det(G \widetilde{G}^{-1}) =\det[I-G (\widetilde{\Sigma}_L-\Sigma_L)]
\end{equation}
where the determinant can be calculated in discretized time slice and real space grid.
In the above equation, we have introduced the following notation:
\begin{equation}  \label{eq55}
\widetilde{G}^{-1}=g^{-1}-\widetilde{\Sigma}_L-\Sigma_R ,\quad
G^{-1}=g^{-1}-\Sigma_L-\Sigma_R  ,
\end{equation}
where $G$ is the Green's function of the quantum dot and $g=g_{nn'}(\tau,\tau')$ denotes the Green's function of the isolated quantum dot, and
\begin{equation}  \label{eq56}
\widetilde{\Sigma}_L(\tau,\tau')
= \Lambda^*(\gamma(\tau))
\left(
  \begin{array}{cc}
    \Sigma^{r}_L  &  \Sigma^{k}_L   \\
     0            &  \Sigma^{a}_L
  \end{array}
\right)   _{(\tau,\tau')}
\Lambda(\gamma(\tau')) ,
\end{equation}
where $\Lambda(\gamma(\tau'))$ and $\Lambda^*(\gamma(\tau))$ is defined in Eq.(\ref{eq48}) and the Green's function and self-energy are written in the Keldysh space in time domain. We can see that the counting field only appears in the self-energy of the left lead in which we count the numbers of the electrons. When $\lambda=\xi=0$, we have $\widetilde{\Sigma}_L=\Sigma_L$.

The Green's function G satisfies the Dyson equation defined on the Keldysh contour from $-\infty$ to $t$ and then back to $-\infty$ with the following relation (for transient regime we should replace $-\infty$ with $0$):
\begin{equation}  \label{eq57}
G(\tau',\tau)=g(\tau',\tau)+\int_{-\infty}^{t}d\tau_1 d\tau_2 g(\tau',\tau_1)\Sigma(\tau_1,\tau_2)G(\tau_2,\tau)  ,
\end{equation}
where $\Sigma(\tau_1,\tau_2)=\Sigma_L(\tau_1,\tau_2)+\Sigma_R(\tau_1,\tau_2)$.
We can write it explicitly as follows
\begin{align}   \label{eq58}
G^{r,a}&=g^{r,a}+g^{r,a}\Sigma^{r,a} G^{r,a}   \notag  \\
G^k&=(1+G^r\Sigma^r)g^k(1+\Sigma^a G^a)+G^r\Sigma^k G^a  .
\end{align}

We point out that if we want to investigate the current correlation between the left and right leads, we should introduce two counting parameters $\lambda_L, \lambda_R$, one for the self-energy of the left lead and another for the right lead, and calculate GF with two counting parameters $Z(\lambda_L,\lambda_R,t)$. For instance, we have $\langle n_L n_R\rangle =\frac{\partial^2 Z(\lambda_L,\lambda_R,t)}{\partial(i\lambda_L)\partial(i\lambda_R)}\big|_{\lambda_L=\lambda_R=0}$. We can also generalize the GF to systems with multiple leads.

The self-energy $\widetilde{\Sigma}_L(\tau,\tau')$ in the presence of the counting field should be calculated separately at four different time regimes. We find from Eqs.(\ref{eq48}) and (\ref{eq56}) that when $-\infty<\tau<0, 0<\tau' <t$, $\widetilde{\Sigma}_L(\tau,\tau')$ is ($\Sigma^{r}_L=0$):
	\begin{equation} \label{eq59}
\widetilde{\Sigma}_L(\tau,\tau')
= e^{-i\xi} \left(
  \begin{array}{cc}
  -i\sin\frac{\lambda}{2}\Sigma^{k}_L   & \cos\frac{\lambda}{2}\Sigma^{k}_L  \\
  -i\sin\frac{\lambda}{2}\Sigma^{a}_L   &   \cos\frac{\lambda}{2}\Sigma^{a}_L
  \end{array}
\right) ,
\end{equation}
and when $0<\tau<t,-\infty<\tau'<0$, we can write $\widetilde{\Sigma}_L(\tau,\tau')$ as ($\Sigma^{a}_L=0$):
\begin{equation} \label{eq60}
\widetilde{\Sigma}_L(\tau,\tau')
= e^{i\xi} \left(
  \begin{array}{cc}
  \cos\frac{\lambda}{2}\Sigma^{r}_L  & \cos\frac{\lambda}{2}\Sigma^{k}_L   \\
 i\sin\frac{\lambda}{2}\Sigma^{r}_L  & i\sin\frac{\lambda}{2}\Sigma^{k}_L
  \end{array}
\right) ,
\end{equation}
and	when $0<\tau,\tau' <t$,
\begin{equation}  \label{eq61}
\widetilde{\Sigma}_L(\tau,\tau')= \exp(i\sigma_x \frac{\lambda}{2}) \Sigma_L(\tau,\tau') \exp(-i\sigma_x \frac{\lambda}{2}) .
\end{equation}
Finally, when $-\infty<\tau, \tau'<0$, we have $\lambda=0$ and $\widetilde{\Sigma}_L(\tau,\tau')=\Sigma_L(\tau,\tau')$. We can see that in transient regime, we only have the case $0<\tau,\tau' <t$, and the parameter $\xi$ does not appear.

Now, we turn to the cumulants of transferred electrons between $t_0=0$ and time $t$ and current of the transient regime. In the transient regime, using the relation $\ln \det \Omega=\mathrm{Tr} \ln \Omega$ we can write the CGF as
\begin{align}   \label{eq62}
\ln Z(\lambda ,t) &=\mathrm{Tr} \ln [I-G (\widetilde{\Sigma}_L-\Sigma_L)]  \notag  \\
&=\mathrm{Tr} \ln [I-G M (e^{-i\sigma_x \lambda}-I)]   ,
\end{align}
where $I$ is the identity matrix and $M$ is given by
\begin{equation} \label{eq63}
M(\tau,\tau')
= \frac{1}{2}\left(
  \begin{array}{cc}
    -\Sigma^a_L +\Sigma^r_L    &   	\Sigma_L^k  \\
    		-\Sigma_L^k   & 	 \Sigma^a_L -\Sigma^r_L   \\
  \end{array}
\right)_{(\tau,\tau')}  .
\end{equation}
Taking the derivative of the CGF with respect to $\lambda$ and using the relation $\mathrm{Tr} \ln (I-\Omega)=-\sum_{j=1}\Omega^j/j$, we can get various cumulants from Eq.~(\ref{eq3}). The first cumulant, the mean number of transferred charge, can be expressed as follows
\begin{align}  \label{eq64}
&\langle\langle \Delta n_L(t) \rangle\rangle  \notag \\
&= \int_{0}^{t}d\tau\int_{0}^{\tau}d\tau'\mathrm{Tr}
[G^r(\tau,\tau')\Sigma_L^<(\tau',\tau)+G^<(\tau,\tau')\Sigma_L^a(\tau',\tau)] \notag   \\
&- \int_{0}^{t}d\tau\int_{0}^{\tau}d\tau'\mathrm{Tr}
[G^a(\tau',\tau)\Sigma_L^<(\tau,\tau')+G^<(\tau',\tau)\Sigma_L^r(\tau,\tau')] ,
\end{align}
which can be written in a more compact form:
\begin{equation} \label{eq65}
\langle\langle \Delta n_L \rangle\rangle
= {\rm Tr}[(G^r-G^a) \Sigma^<_L + G^< (\Sigma^a_L-\Sigma^r_L)]
\end{equation}
where the trace is over both time space and real space. Similarly, the charge-charge correlation (the second cumulant) is found to be\cite{gm}
\begin{equation} \label{eq66}
\langle\langle (\Delta n_L)^2 \rangle \rangle= -{\rm Tr}[(G M \sigma_x)^2+GM].
\end{equation}
From $\langle\langle \Delta n_L(t) \rangle\rangle=\int_0^t I_L(\tau) d\tau$, we find the current at time $t$,
\begin{align} \label{eq67}
I_L(t) = \int_{0}^{t}d\tau {\rm Tr}[G^r(t,\tau)\Sigma_L^<(\tau,t)+G^<(t,\tau)\Sigma_L^a(\tau,t)]+H.c..
\end{align}
The current here is quite different from Cini's approach (the partition free approach), where the coupling between leads and the central quantum dot is turned on in the infinite past while the bias is applied at $t_0=0$. \cite{28,29,Cini} In our approach, both the coupling and the bias are turned on at $t_0=0$. 


It is not difficult to prove that we obtain exactly the same expression for the average current as in Eq.(\ref{eq67}) in the measurement regime where two measurements are performed in the dc case. However, the second and higher cumulants in the measurement regime are not the same as those of the transient regime. This confirms the fact that the first measurement does perturb the system and therefore the current under dc bias is not a constant after the measurement. Similar behavior has been found previously in the case of phonon transport.\cite{JS1,JS2}


We can derive the long-time behavior of the generating function which recovers the famous Levitov-Lesovik formula.\cite{Levitov2,Levitov3,Klich,Levitov4}. This has been discussed in detail in the papers of M.~Esposito \textit{et al.} \cite{RMP} and Agarwalla \textit{et al.}.\cite{JS2} For completeness of this paper, we just give a brief summary here about how to get the long-time limit from the FCS in the transient regime. For convenience we assume that we switch on the interaction between the subsystems at $-t/2$ and we are interested in the behavior between time $-t/2$ and $t/2$. When $t\rightarrow \infty$, the interval becomes $(-\infty, \infty)$, and the Green's function and the self-energy in the time domain are invariant under the time translation. The CGF, the logarithm of the determinant of Eq.~(\ref{eq54}), in the energy space in the long-time limit is
\begin{equation}  \label{long1}
\ln Z_s(\lambda ,t)=t\int \frac{d\omega}{2\pi} \ln\det\left\{ 1-G(\omega)[\widetilde{\Sigma}_L(\omega)-\Sigma_L(\omega)] \right\}  .
\end{equation}
If we use the first equality of the determinant of Eq.~(\ref{eq54}), CGF in the energy space in the long-time limit can be expressed as,
\begin{align}    \label{long2}
&\ln Z_s(\lambda ,t)=t\int \frac{d\omega}{2\pi}\ln\det\left[ \left( \begin{array}{cc}
G^r  &  0  \\   0   &   G^a
\end{array} \right) \times \right.   \notag  \\
&\left. \left( \begin{array}{cc}
(g^r)^{-1}-\widetilde{\Sigma}_{L}^{11}-\Sigma_{R}^r
&  -\widetilde{\Sigma}_L^{12}-\Sigma_{R}^k   \\
-\widetilde{\Sigma}_L^{21}   &  (g^a)^{-1}-\widetilde{\Sigma}_{L}^{22}-\Sigma_{R}^a
\end{array} \right) \right]   ,
\end{align}
where we ignore $G^k$ in the determinant since the Green's function is an upper-triangle block matrix in the Keldysh space, and $\widetilde{\Sigma}_{L}^{11}$. 
Using the relations $(G^r)^{-1}-(G^a)^{-1}=\Sigma^a-\Sigma^r$ and Eq.(\ref{eqnew}), we have
\begin{align}    \label{long3}
\ln Z_s(\lambda ,t)=& t\int \frac{d\omega}{2\pi}\ln\det [ I+G^r(\Sigma_R^r-\Sigma_R^a)G^a(e^{i\lambda}-1)\Sigma_L^<   \notag   \\
&+G^r\Sigma_R^< G^a(e^{-i\lambda}-1)(\Sigma_L^r-\Sigma_L^a)  \notag   \\
&+G^r\Sigma_R^<G^a(e^{i\lambda}+e^{-i\lambda}-2)\Sigma_L^<   ]
\end{align}
Further using the relations $\Sigma_{\alpha}^r-\Sigma_{\alpha}^a =-i\Gamma_{\alpha}$ and $\Sigma_{\alpha}^<=i\Gamma_{\alpha}f_{\alpha}$, we obtain the CGF in the long-time limit as
\begin{align}  \label{long4}
\ln Z_s(\lambda)=& t\int \frac{d\omega}{2\pi}{\rm Tr} \ln \{ I+T(\omega)[(e^{i\lambda}-1)(1-f_R(\omega))f_L(\omega)   \notag  \\
& +(e^{-i\lambda}-1)(1-f_L(\omega))f_R(\omega)] \}
\end{align}
with the transmission coefficient for the quantum dot $T(\omega)=G^r\Gamma_L G^a\Gamma_R$. Next we get the current generating function $S_s(\lambda)$
\begin{align}  \label{long5}
S_s(\lambda)=&\lim_{t\rightarrow \infty}\frac{\ln Z_s(\lambda)}{t}    \notag  \\
=& \int \frac{d\omega}{2\pi}{\rm Tr}\ln \left\{ I+T(\omega)[(e^{i\lambda}-1)(1-f_R(\omega))f_L(\omega)   \right. \notag  \\
& \left. +(e^{-i\lambda}-1)(1-f_L(\omega))f_R(\omega)] \right\}
\end{align}
which is the celebrated Levitov-Lesovik formula. Taking the derivative of the current generating function with respect to $\lambda$ at $\lambda=0$, we get the current of the steady state in the long-time limit which is the Landauer-Buttiker formula.\cite{L-B}
\begin{equation}  \label{long6}
I(t)=\int \frac{d\omega}{2\pi}T(\omega)[f_L(\omega)-f_R(\omega)]  .
\end{equation}

Finally, we wish to emphasize that the formalism discussed here cannot be used to study the short time full-counting statistics in dc steady state quantum transport since the first measurement is not non-invasive. A formalism of short time FCS in dc steady state within nonequilibrium Green's function formalism is still unknown.

\section{Generalization to Magnetic Tunnel Junction}

\begin{figure}
  \includegraphics[width=3.0in]{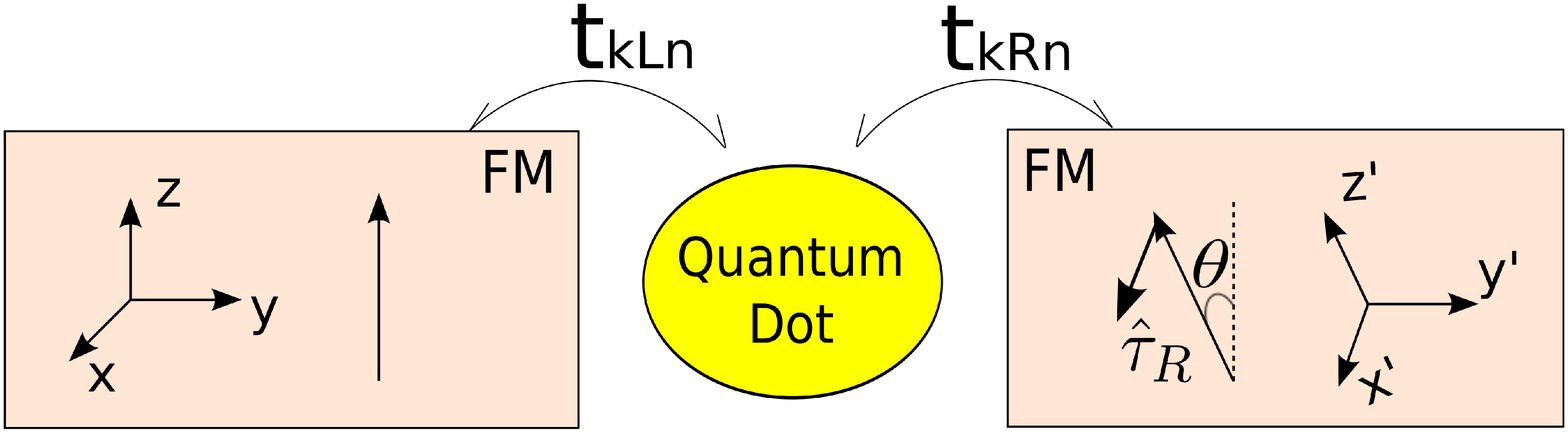}\\
  \caption{Schematic diagram of a magnetic tunnel junction (MTJ) in which the left and right ferromagnetic leads are coupled to the non-magnetic scattering region which is a quantum dot in our case. The magnetic moment $\mathbf{M}$ of the left lead is along the $z-$axis, while the magnetic moment of the right lead is at an angle of $\theta$ to the $z-$axis, which is along the $z'$ axis. }
  \label{fig3}
\end{figure}

In this section, we generalize the formalism discussed above to FCS in spintronics. As an example we study a magnetic tunnel junction (MTJ) in which the left and right ferromagnetic leads are coupled to the non-magnetic scattering region which is a quantum dot. The magnetic moment $\mathbf{M}$ of the left lead is along the $z-$axis, while the magnetic moment of the right lead is at an angle of $\theta$ to the $z-$axis, which is along the $z'$ axis (such that the coordinate system $x' y' z'$ is obtained by rotating the coordinate system $xyz$ by an angle $\theta$ along the $y$ direction), the electric current flows in the $y-$direction (see Fig.{\ref{fig3}}). The relative orientation of the magnetizations (parallel or anti-parallel) in the two electrodes will induce the tunnel magnetoresistance (TMR) effect. \cite{TMR1,TMR2,TMR3}  The magnetization switching probability by non-Gaussian spin-torque shot noise is recently studied by taking FCS into consideration and using the fluctuation theorem. \cite{switching} Here, we present a formalism using NEGF which is suitable to study the FCS of transient behaviors in MTJ. Treating $xyz$ coordinate system as the frame of reference, the Hamiltonian of the whole system reads as
\begin{equation}  \label{mtj1}
H=H_L+H_R+H_{dot}+H_{T}
\end{equation}
where $H_L$ and $H_R$ describe the Hamiltonian of the left and right leads:
\begin{align}   \label{mtj2}
H_L=&\sum_{kL\sigma}(\epsilon_{kL}-\sigma M_L )C_{kL\sigma}^{\dag}C_{kL\sigma}  \notag \\
H_R=&\sum_{kR\sigma}(\epsilon_{kR}-\sigma M_R\cos\theta)C_{kR\sigma}^{\dag}C_{kR\sigma} \notag \\
&- M_R\sin\theta C_{kR\sigma}^{\dag}C_{kR\bar{\sigma}}   ,
\end{align}
$H_{dot}$ describes the non-magnetic scattering region (quantum dot),
\begin{equation}   \label{mtj3}
H_{dot}=\sum_{n\sigma} \epsilon_n C_{n\sigma}^{\dag}C_{n\sigma} ,
\end{equation}
$H_T$ is the Hamiltonian that models the coupling between leads and the quantum dot with hopping matrix.
\begin{align}   \label{mtj4}
H_T&=H_{LS\uparrow}+H_{LS\downarrow}+H_{RS\uparrow}+H_{RS\downarrow}  \notag  \\
&=\sum_{k\alpha n\sigma}[t_{k\alpha n} C_{k\alpha\sigma}^{\dag}C_{n\sigma}+H.c.] .
\end{align}
In these representations, $\alpha$ represents $L$ or $R$, $C_{k\alpha \sigma}^{\dag}$ (with $\sigma =\uparrow, \ \downarrow$ or $\pm 1$, and $\bar{\sigma}=-\sigma$) is the creation operator of electrons at energy level $k$ with spin index $\sigma$ inside the left or right lead. Similarly, $C_{n\sigma}^{\dag}$ is the creation operator of electrons at energy level $n$ with spin index $\sigma$ inside the quantum dot.

To diagonalize the Hamiltonian of the right lead, we apply the following Bogoliubov transformation to the creation and annihilation operator of the right lead, \cite{Bogoliubov}
\begin{equation}   \label{mtj5}
C_{kR}=R \ c_{kR} ,\qquad  C_{kR}^\dag= c_{kR}^{\dag} \ R^\dag
\end{equation}
where we have used the abbreviation $C_{k\alpha}=\left(
\begin{array}{c}
C_{kR\uparrow} \\  C_{kR\downarrow}
\end{array}
\right) , c_{k\alpha}=\left(
\begin{array}{c}
c_{kR\uparrow} \\  c_{kR\downarrow}
\end{array}
\right)$, and $R=\left(
\begin{array}{cc}
\cos \frac{\theta}{2} & -\sin \frac{\theta}{2}   \\
\sin \frac{\theta}{2}  & \cos \frac{\theta}{2}
\end{array}
\right)$ in the transformation, while the creation and annihilation operators of the left lead and central quantum dot remain unchanged, then we can get the effective Hamiltonian
\begin{align}   \label{mtj6}
&H_{\alpha} =
\sum_{k\alpha} (\epsilon_{k\alpha}- \sigma M) c_{k\alpha\sigma}^{\dag} c_{k\alpha\sigma} \notag \\
&H_{dot} = \sum_{n} \epsilon_n c_n^{\dag} c_n   \notag  \\
&H_{T} =\sum_{k\alpha n}(c_{k\alpha}^{\dag} t_{k\alpha n} R^{\dag} c_n +H.c.)
=\sum_{k\alpha n}(c_{k\alpha}^{\dag}\mathbf{T}_{k\alpha n} c_n +H.c.)
\end{align}
where we used the abbreviation $c_{k\alpha}=\left(
\begin{array}{c}
c_{k\alpha\uparrow} \\  c_{k\alpha\downarrow}
\end{array}
\right)$, $c_n=\left(
\begin{array}{c}
c_{n\uparrow} \\  c_{n\downarrow}
\end{array}
\right)$, $\mathbf{T}_{kLn}=t_{kLn}$ and $\mathbf{T}_{kRn}=t_{kR n} R^{\dag}$ in the expression of $H_T$. From now on, we use capital cases $C_{k\alpha}^{\dag}$, $C_{k\alpha}$, $C_{n}^{\dag}$, $C_{n}$ to denote the creation and annihilation operators of the leads and quantum dot before Bogoliubov transformation while use $c_{k\alpha}^{\dag}$, $c_{k\alpha}$, $c_{n}^{\dag}$, $c_{n}$ to denote the creation and annihilation operators after the transformation.

\bigskip

{\noindent \bf FCS of transferred charge with a particular spin direction}

\bigskip

Now we count the number of electrons with spin-up and spin-down in $z$ direction in the left lead separately under the transient regime. For the right lead, we count the number of electrons with spin-up and spin-down in $z'$ direction.  For convenience, we just consider the spin-up case and the case for spin-down is self-evident. As was demonstrated in the last section that the counting field just enters the coupling term between quantum dot and the particular lead so the modified Hamiltonian $H_{\gamma}$ with regard to the spin-up number operator $\hat{N}_{\alpha\uparrow}^{(h)}(t)=\sum_{k} c_{k\alpha\uparrow}^{\dag}(t)c_{k\alpha\uparrow}(t)$ can be written as follows,
\begin{align} \label{mtj7}
H_{\gamma}(t)&=e^{i\gamma N_{\alpha\uparrow}(0)}H(t)e^{-i\gamma N_{\alpha\uparrow}(0)}  \notag \\
&=H_{lead}+H_{dot}+H_{\bar{\alpha}S}
+\mathrm{diag}(e^{i\gamma} , 1)\sum_{k} c_{k\alpha}^{\dag} \mathbf{T}_{k\alpha n} c_{n} \notag  \\
&+\mathrm{diag}(e^{-i\gamma},1)\sum_{k} c_{n}^{\dag} \mathbf{T}_{nk\alpha}c_{k\alpha},
\end{align}
where $\bar{\alpha}=R$, if $\alpha=L$ and vice versa, $\mathbf{T}_{nkL}=t_{nkL}$, $\mathbf{T}_{nkR}=R t_{nkR}$.
If we are working in the $xyz$ coordinate system for the left lead, we have the modified self-energy of the left lead
\begin{equation}  \label{mtj8}
\widetilde{\Sigma}_L(\tau,\tau')=\left(
  \begin{array}{cc}
\Lambda^* \Sigma_{L\uparrow}  \Lambda  & 0  \\
0   &   \Sigma_{L\downarrow}
  \end{array}
\right)
\end{equation}
where $\Sigma_{\alpha\sigma}$ is defined as
$\Sigma_{\alpha\sigma}(\tau,\tau') =
\sum_{k,k'}t_{nk\alpha} g_{kk'\alpha\sigma}(\tau,\tau')t_{k'\alpha n'} $, $\Lambda^*$ and $\Lambda$ act on the Keldysh space. $\Lambda$ is almost the same as Eq.(\ref{eq48}),
\begin{equation} \label{mtj9}
\Lambda =\exp\left(-\frac{i\lambda}{2} \sigma_x  \right)
= \left(
  \begin{array}{cc}
    \cos\frac{\lambda}{2} & -i\sin\frac{\lambda}{2}   \\
    -i\sin\frac{\lambda}{2} & \cos\frac{\lambda}{2}
  \end{array}  \right) ,
\end{equation}
due to the fact that in the transient regime, $\Lambda$ doesn't depend on time and the parameter $\xi$ disappears. The normalized GF can be written as:
\begin{equation}  \label{mtj10}
Z_{L\uparrow}(\lambda,t)=\det (G \widetilde{G}^{-1})
\end{equation}
where
\begin{align}  \label{mtj11}
\widetilde{G}^{-1}&=g^{-1}-\widetilde{\Sigma}_L-R\Sigma_R R^\dag  \notag \\
G^{-1}&=g^{-1} -\Sigma_L -R \Sigma_R R^\dag .
\end{align}
The Green's function $g$ is for the diagonalized Hamiltonian of the central quantum dot, $R$ and $R^\dag$ act on the spin space. Similar expression of GF can be obtained for spin-down electrons of the left lead by modification of Eq.(\ref{mtj8}).

Similar to case of the left lead, if we count the number of electrons with spin-up in $z'$ direction $\sum_{k}c_{kR\uparrow}^\dag c_{kR\uparrow} $ in the right lead, the normalized GF can be written as:
\begin{equation}  \label{mtj12}
Z_{R\uparrow}(\lambda,t)=\det (G \widetilde{G}^{-1}) , 
\end{equation}
where
\begin{align}   \label{mtj13}
\widetilde{G}^{-1}&=g^{-1}-\Sigma_L-R \widetilde{\Sigma}_R R^\dag ,   \notag \\
G^{-1}&=g^{-1} -\Sigma_L -R \Sigma_R R^\dag
\end{align}
with
\begin{equation}  \label{mtj14}
\widetilde{\Sigma}_R(\tau,\tau')=\left(
  \begin{array}{cc}
\Lambda^* \Sigma_{R\uparrow}  \Lambda  & 0  \\
0   &   \Sigma_{R\downarrow}
  \end{array}
\right) .
\end{equation}

Here we point out that the GF for the spin-up electrons $\sum_{k}C_{kR\uparrow}^\dag C_{kR\uparrow}$ in the $z$ direction of the right lead is totally different from that of $z'$. For the spin-up electrons $\sum_{k}C_{kR\uparrow}^\dag C_{kR\uparrow}$ in the $z$ direction of the right lead, the corresponding modified Hamiltonian is
\begin{align*}
H_{\gamma}(t)
=H_{lead}+H_{dot}+H_{LS}
+\left( \sum_{k n}C_{kR}^{\dag} \tilde{t}_{kR n} C_n  +H.c. \right)
\end{align*}
with $\tilde{t}_{kR n}=\left( \begin{array}{cc}
e^{i\gamma}t_{kRn}  &  0  \\   0   &   t_{kRn}
\end{array}  \right) $. After Keldysh rotation, $\tilde{t}_{kR n}$ becomes $\bar{t}_{kR n}=\left( \begin{array}{cc}
t_{kRn} \Lambda  &  0  \\   0   &   t_{kRn}
\end{array}  \right) $.
Because of this we have
\begin{align*}
&\widetilde{G}^{-1}
= g^{-1}-\Sigma_L-
\sum_{kk'}\bar{t}_{nkR}
     R \  g_{kk'}  \ R^{\dag}
\bar{t}_{k'R n}   \notag \\
&= g^{-1}-\Sigma_L  \notag \\
&-\left(
  \begin{array}{cc}
\Lambda^* (\cos^2\frac{\theta}{2}\Sigma_{R\uparrow}+\sin^2\frac{\theta}{2}\Sigma_{R\downarrow})  \Lambda  &
\frac{1}{2}\sin\theta \Lambda^* (\Sigma_{R\uparrow}-\Sigma_{R\downarrow})  \\
\frac{1}{2}\sin\theta (\Sigma_{R\uparrow}-\Sigma_{R\downarrow})  \Lambda  &
\sin^2\frac{\theta}{2}\Sigma_{R\uparrow}+\cos^2\frac{\theta}{2}\Sigma_{R\downarrow}
  \end{array}
\right)
\end{align*}
where we have used the short notation $g_{kk'}=\left(
\begin{array}{cc}
g_{kk'R\uparrow}  &  0   \\  0   &  g_{kk'R\downarrow}
\end{array}  \right) $.

\bigskip

{\noindent \bf FCS of transferred charge current and spin current}

\bigskip

We know that the total charge current operator through lead $\alpha$ is
\begin{equation}   \label{mtj15}
\hat{I}_{\alpha}=\hat{I}_{\alpha\uparrow}+\hat{I}_{\alpha\downarrow} ,
\end{equation}
while the spin current operator should be
\begin{equation}  \label{mtj16}
\hat{I}_{\alpha}^s=\frac{\hbar}{2q}(\hat{I}_{\alpha\uparrow}-\hat{I}_{\alpha\downarrow}) ,
\end{equation}
with $\hat{I}_{\alpha\sigma}=q\frac{d\hat{N}_{\alpha\sigma}}{dt}$, $\hat{N}_{\alpha\sigma}=\sum_k \hat{c}_{k\alpha\sigma}^\dag \hat{c}_{k\alpha\sigma}$
and we can set $\hbar=q=1$ here. The modified self-energy in the GF of the number of total charge transferred in the lead $\alpha$ is (when $\alpha=L (R)$ we consider $z$ ($z'$)direction)
\begin{equation}   \label{mtj17}
\widetilde{\Sigma}_\alpha (\tau,\tau')=\left(
  \begin{array}{cc}
\Lambda^* \Sigma_{\alpha\uparrow}  \Lambda  & 0  \\
0   &   \Lambda^* \Sigma_{\alpha\downarrow} \Lambda
  \end{array}
\right)
\end{equation}
and modified self-energy in the GF of the total spin transferred
\begin{equation}   \label{mtj18}
\widetilde{\Sigma}_\alpha(\tau,\tau')=\left(
  \begin{array}{cc}
\bar{\Lambda}^* \Sigma_{\alpha\uparrow}  \bar{\Lambda}   & 0  \\
0   &   \bar{\Lambda} \Sigma_{\alpha\downarrow} \bar{\Lambda}^*
  \end{array}
\right)
\end{equation}
with short notation $\bar{\Lambda} =\exp\left(-\frac{i\lambda}{4} \sigma_x  \right) $.\cite{foot2}

Note that GF for the total transferred charge (or total transferred spin) $Z\neq Z_{\alpha\uparrow} Z_{\alpha\downarrow}$ since the statistics for spin-up and spin-down transferred electrons are not independent of each other because of the presence of spin flip mechanism. Hence we cannot directly use the GF for the spin-up and spin-down to obtain the GF for the statistics of the total transferred charge or spin.  Similar to Eq.(\ref{long3}), the CGF in the long-time limit in the energy space for the number of total transferred charge or spin in the right lead can be expressed as
\begin{align}    \label{long7}
\ln Z_s(\lambda ,t)=& t\int \frac{d\omega}{2\pi}\ln\det [ I  \notag   \\
&+G^r(\Sigma_L^r-\Sigma_L^a)G^a R \Upsilon \Sigma_R^< R^\dag  \notag   \\
&+G^r\Sigma_L^< G^a R \Upsilon^\dag (\Sigma_R^r-\Sigma_R^a) R^\dag \notag   \\
&+G^r\Sigma_L^<G^a R (\Upsilon+\Upsilon^\dag)\Sigma_R^< R^\dag  ]
\end{align}
where for the total transferred charge we take $\Upsilon=\mathrm{diag}(e^{i\lambda}-1, e^{i\lambda}-1)$ while for the total transferred spin we take $\Upsilon=\mathrm{diag}(e^{i\lambda/2}-1, e^{-i\lambda/2}-1)$.

\bigskip

{\noindent \bf FCS of spin-transfer torque}

\bigskip

The total spin torque operator can be derived from the total spin along the $x'$ direction in the right ferromagnetic electrode, \cite{GangSu, YunjinYu,foot18}
\begin{equation}  \label{mtj19}
\hat{S}_{x'}=\frac{\hbar}{2}\sum_{k} c_{kR}^{\dag} \sigma_x c_{kR}
=\frac{\hbar}{2}
\sum_{k}\left(c_{kR\uparrow}^{\dag}c_{kR\downarrow}+c_{kR\downarrow}^{\dag}c_{kR\uparrow}
\right) .
\end{equation}
The spin transfer torque operator is
\begin{align}   \label{mtj20}
\hat{\tau}_R = \frac{i}{\hbar}[\hat{H},\hat{S}_{x'}]
= -\frac{i}{2} \sum_{k} t_{kRn} c_{kR}^\dag \bar{R} c_n +H.c.
\end{align}
with $\bar{R}=\left(
\begin{array}{cc}
-\sin \frac{\theta}{2} &  \cos \frac{\theta}{2}  \\
 \cos \frac{\theta}{2} &  \sin \frac{\theta}{2}
\end{array}
\right)$.
Now, we can write the modified Hamiltonian $H_{\gamma}$ with regard to the total spin operator $\hat{S}_{x'}$ using the Baker-Hausdorff lemma Eq.(\ref{baker})($\hbar=1$),
\begin{align}   \label{mtj21}
H_{\gamma}(t)
&=e^{i\gamma \hat{S}_{x^{\prime}}(0)}H(t)e^{-i\gamma \hat{S}_{x^{\prime}}(0)}  \notag \\
&=H_{lead}+H_{dot}+H_{LS} \notag  \\
&+\left\{ \frac{e^{i\gamma/2}+e^{-i\gamma/2}}{2} \sum_{kRn}c_{kR}^\dag t_{kRn}R^\dag c_n +H.c. \right\} \notag \\
&+\left\{ \frac{e^{i\gamma/2}-e^{-i\gamma/2}}{2} \sum_{kRn}c_{kR}^\dag t_{kRn}\bar{R} c_n +H.c. \right\}  .
\end{align}
Comparing with Eq.(\ref{eq29}) for the case of number of transferred charges, we can easily write the normalized GF for the total spin (whose time derivative is spin transfer torque) as follows:
\begin{equation}   \label{mtj22}
Z_{x'}(\lambda,t)=\det (G \widetilde{G}^{-1})
\end{equation}
where
\begin{align}   \label{mtj23}
\widetilde{G}^{-1}=g^{-1}&-\Sigma_L   \notag  \\
&-R \widetilde{\Sigma}_{R1} R^\dag
-R \widetilde{\Sigma}_{R2} \bar{R}
-\bar{R} \widetilde{\Sigma}_{R3} R^\dag
-\bar{R} \widetilde{\Sigma}_{R4} \bar{R}
  \notag \\
G^{-1}=g^{-1} &-\Sigma_L
-R \Sigma_R R^\dag
\end{align}
with
\begin{align}    \label{mtj24}
\widetilde{\Sigma}_{R1}(\tau,\tau') &=\left(
  \begin{array}{cc}
\Xi_1^* \Sigma_{R\uparrow}  \Xi_1  & 0  \\
0   &   \Xi_1^* \Sigma_{R\downarrow} \Xi_1
  \end{array}
\right) \notag  \\
\widetilde{\Sigma}_{R2}(\tau,\tau') &=\left(
  \begin{array}{cc}
\Xi_1^* \Sigma_{R\uparrow} \Xi_2  & 0  \\
0   &   \Xi_1^* \Sigma_{R\downarrow} \Xi_2
  \end{array}
\right) \notag  \\
\widetilde{\Sigma}_{R3}(\tau,\tau') &=\left(
  \begin{array}{cc}
\Xi_2^* \Sigma_{R\uparrow}  \Xi_1  & 0  \\
0   &   \Xi_2^* \Sigma_{R\downarrow} \Xi_1
  \end{array}
\right) \notag  \\
\widetilde{\Sigma}_{R4}(\tau,\tau') &=\left(
  \begin{array}{cc}
\Xi_2^* \Sigma_{R\uparrow}  \Xi_2  & 0  \\
0   &   \Xi_2^* \Sigma_{R\downarrow} \Xi_2
  \end{array}
\right)
\end{align}
and
\begin{align}   \label{mtj25}
\Xi_1 &= \frac{\bar{\Lambda} + \bar{\Lambda}^* }{2}
=\left(
  \begin{array}{cc}
\cos\frac{\lambda}{4}   & 0  \\
0   &   \cos\frac{\lambda}{4}
  \end{array}
\right)  \notag  \\
\Xi_2 &= \frac{\bar{\Lambda} - \bar{\Lambda}^* }{2}
=\left(
  \begin{array}{cc}
0   &   -i\sin\frac{\lambda}{4}  \\
 -i\sin\frac{\lambda}{4}  &  0
  \end{array}
\right)  .
\end{align}
Here, we point out that $\Xi_1,\ \Xi_1^* , \ \Xi_2, \ \Xi_2^*$ act on the Keldysh space while $R, \ R^\dag, \ \bar{R}$ act on the spin space of self-energy in the GF. Note that $\bar{R}=\left(\begin{array}{cc} 0 & 1 \\ 1 & 0 \end{array}\right)R^\dag$; we can rewrite $ \widetilde{G}^{-1}$ in Eq.(\ref{mtj23}) in the following form:
\begin{equation} \label{mtj26}
\widetilde{G}^{-1}=g^{-1}-\Sigma_L -R \widetilde{\Sigma}_R R^\dag
\end{equation}
with
\begin{equation}  \label{mtj27}
\widetilde{\Sigma}_R =
\left(  \begin{array}{cc}
\Xi_1^*  & \Xi_2^*  \\   \Xi_2^*  &   \Xi_1^*
\end{array}  \right)
\left( \begin{array}{cc}
 \Sigma_{R\uparrow}    & 0  \\  0   &   \Sigma_{R\downarrow}
\end{array}  \right)
\left(  \begin{array}{cc}
\Xi_1  & \Xi_2  \\   \Xi_2  &   \Xi_1
\end{array}  \right) .
\end{equation}
	
Just like Eq.~(\ref{long1}), we can get the expression of CGF of the spin-transfer torque in the energy space in the long-time limit as
\begin{align}  \label{mtj28}
&\ln Z_s(\lambda ,t)  \notag  \\
&=t\int \frac{d\omega}{2\pi}{\rm Tr}\ln\det[1-G(\omega)R(\widetilde{\Sigma}_R(\omega)-\Sigma_R(\omega))R^{\dag}] .
\end{align}
It can be easily shown that the spin-transfer torque from this equation is the same as that derived from Ref.\onlinecite{GangSu}.

\section{Quantum Point Contact}

In this section, we extend the formalism further to the quantum point contact (QPC) system which is the simplest in mesoscopic systems and its transport properties have been studied extensively. The difference between the QPC and the quantum dot system studied in the previous sections is that in QPC, two electrodes are connected directly by the hopping term; this is experimentally achieved by a narrow constriction between the electrodes. Examples of two electrodes involved are conductor-superconductor (N-S) and superconductor-superconductor (S-S) systems. \cite{Cuevas}
Such a system can be described by the following simple Hamiltonian:
\begin{equation} \label{qpc1}
H=H_0+H_T=H_{L}+H_{R}+H_T
\end{equation}
where $H_0$ consists of the Hamiltonian of the isolated electrodes,
\begin{equation}  \label{qpc2}
H_0=\sum_{x\in k\alpha} \epsilon_{x} c_{x}^{\dag}c_{x},
\end{equation}
where we use the index $k\alpha$ to label the states of the electrode $\alpha$. Here, $\epsilon_{k\alpha}=\epsilon_{k\alpha}^{(0)}+q\Delta_{\alpha}(t)$, where $\epsilon_{k\alpha}^{(0)}$ is the energy levels in electrode $\alpha$ and $\Delta_{\alpha}(t)$ is the external voltage, and $H_T$ is the Hamiltonian describing the direct hopping between the nearest-neighbor sites in the two electrodes with a coupling constant $t_{LR}=t_{RL}^*$:
\begin{equation}
H_T= t_{LR} c_{L}^{\dag}c_{R}+t_{RL}c_{R}^{\dag}c_{L}  .
\end{equation}

We count the number of transferred electrons in the left electrode, and the electrons flow from the left electrode to the right one is defined as positive direction of the current. Following the discussion of the quantum dot system in Sec. III,
in accordance with Eqs.(\ref{eq51}) and (\ref{eq52}) we can express the GF as follows:
\begin{equation}  \label{qpc3}
Z(\lambda,t)=\frac{\det \mathcal{M}(\lambda)}{\det \mathcal{M}(\lambda=0)}
\end{equation}
with
\begin{equation} \label{qpc4}
\mathcal{M}=
\left(
  \begin{array}{cc}
    g_{L L}^{-1}(\tau,\tau') & -t_{LR}(\tau,\tau)\Lambda   \\
  -\Lambda^* t_{RL}(\tau,\tau) &  g_{RR}^{-1}(\tau,\tau')
  \end{array}
\right),
\end{equation}
where $\Lambda$ is the same as Eq.(\ref{mtj9}) for the transient regime. For convenience, we introduce the following abbreviated notation:
\begin{equation}  \label{qpc6}
\mathbf{\widetilde{G}}^{-1}=\textbf{g}^{-1}-\mathbf{\tilde{t}}, \qquad
\mathbf{G}^{-1}=\textbf{g}^{-1}-\mathbf{t}
\end{equation}
with
\begin{align}   \label{qpc7}
	\mathbf{t}=
	\left(  \begin{array}{cc}
     & t_{LR}   \\   t_{RL}  &
  	\end{array}  \right) ,&\ \
	\mathbf{\tilde{t}}=
	\left(  \begin{array}{cc}
     & t_{LR}\Lambda   \\   \Lambda^* t_{RL}  &
  	\end{array}  \right)  ,\notag  \\
  	\textbf{g}^{-1}=&
	\left( \begin{array}{cc}
     g_{LL}^{-1} &    \\    &  g_{RR}^{-1}
  	\end{array}  \right)  .
\end{align}
As mentioned in Sec. III, $g_{LL}^{-1}$ and $g_{RR}^{-1}$ contain the Keldysh components and $t_{LR}$ and $t_{RL}$ are diagonal matrices in Keldysh space. Then, we write GF as
\begin{equation}  \label{qpc8}
Z(\lambda,t)=\det(\mathbf{G} \mathbf{\widetilde{G}}^{-1} )
=\det[\mathbf{I}-\mathbf{G}(\mathbf{\tilde{t}}-\mathbf{t})]  .
\end{equation}
For Green's function, we have the following Dyson equation in Keldysh space
\begin{equation}
\mathbf{G}=\mathbf{g}+\mathbf{g}\mathbf{t}\mathbf{G}  .
\end{equation}
We can write the Dyson equation explicitly as \cite{Cuevas}
\begin{align}  \label{qpc9}
\mathbf{G}^{r,a}=\mathbf{g}^{r,a}+\mathbf{g}^{r,a}\mathbf{t}^{r,a}\mathbf{G}^{r,a} ,
\notag  \\
\mathbf{G}^{k}
=(\mathbf{I}+\mathbf{G}^{r}\mathbf{t}^{r})\mathbf{g}^{k}(\mathbf{I}+\mathbf{t}^{a}\mathbf{G}^{a}) , \notag  \\
\mathbf{G}^{<}
=(\mathbf{I}+\mathbf{G}^{r}\mathbf{t}^{r})\mathbf{g}^{<}(\mathbf{I}+\mathbf{t}^{a}\mathbf{G}^{a})
\end{align}
with
\begin{align}  \label{qpc10}
\mathbf{G}^{r,a,k} &=	
\left(  \begin{array}{cc}
     G_{LL}^{r,a,k}  & G_{LR}^{r,a,k}   \\   G_{RL}^{r,a,k}   &  G_{RR}^{r,a,k}
\end{array}  \right)    ,\quad
\mathbf{g}^{r,a,k} =	
\left(  \begin{array}{cc}
     g_{LL}^{r,a,k}  & 0   \\   0   &  g_{RR}^{r,a,k}
\end{array}  \right)    ,\qquad  \notag  \\
\mathbf{t}^{r,a} &=
\left(  \begin{array}{cc}
     t_{LL}^{r,a}  & t_{LR}^{r,a}   \\  t_{RL}^{r,a}  & t_{RR}^{r,a}
\end{array}  \right) =
\left(  \begin{array}{cc}
      0 &  t_{LR}    \\  t_{RL}   & 0
\end{array}  \right)  ,
\end{align}
and $\mathbf{t}^{k}=0$ as previously mentioned that $\mathbf{t}$ is diagonal in Keldysh space.

Now, we turn to the cumulants of transferred electrons between $t_0=0$ and time $t$ and current of transient regime. In the transient regime, from the fact $\ln \det \Omega=\mathrm{Tr} \ln \Omega$ we can write the CGF as
\begin{equation}  \label{qpc11}
\ln Z(\lambda ,t) =\mathrm{Tr} \ln [\mathbf{I}-\mathbf{G}(\mathbf{\tilde{t}}-\mathbf{t})]  .
\end{equation}
Taking the derivative of the CGF with respect to $\lambda$ and using the relation $\mathrm{Tr} \ln (I-\Omega)=-\sum_{j=1}\Omega^j/j$, we can get various cumulants from Eq.~(\ref{eq3}). Using the relations ${\rm Tr}\ t_{LR}G_{RL}^r ={\rm Tr}\ t_{RL}G_{LR}^r $ and $G^k=2G^<+G^r-G^a$, the first cumulant, the mean number of transferred charge, can be expressed as,
\begin{align} \label{qpc12}
\langle\langle \Delta n_L \rangle\rangle
&= {\rm Tr}\left[-\mathbf{G}\frac{\partial\mathbf{\tilde{t}}}{\partial(i\lambda)}   \right]\bigg|_{\lambda=0}   \notag  \\
&={\rm Tr} \left( \frac{1}{2}t_{LR}G_{RL}^k -\frac{1}{2}t_{RL}G_{LR}^k \right)  \notag  \\
&=\int_0^{t}d\tau \left( t_{LR}G_{RL}^<(\tau,\tau) -t_{RL}G_{LR}^<(\tau,\tau) \right) ,
\end{align}
Hence, from $\langle\langle \Delta n_L(t) \rangle\rangle=\int_0^t I_L(\tau) d\tau$, we can get the transient current at time $t$:
\begin{align} \label{qpc13}
I_L(t) =   t_{LR}G_{RL}^<(t,t) -t_{RL}G_{LR}^<(t,t)  .
\end{align}
We note that a similar expression has been obtained in the dc case.\cite{Cuevas} We point out that the derivation above can be easily generalized to a QPC system with multiple electrodes, or the systems with spin configuration such as N-S or S-S system.

\section{Numerical Results}

We now apply the formalism discussed above to a system in which two single-level quantum dots are in series and connected to the left and right leads respectively. The Hamiltonian of such a system reads as
\begin{equation}  \label{nr1}
H_0=\epsilon_1 d_1^\dag d_1 + \epsilon_2 d_2^\dag d_2 + t_{12}d_1^\dag d_2 +t_{21} d_2^\dag d_1, 
\end{equation}
where $\epsilon_1$ and $\epsilon_2$ are the two energy levels of the quantum dots and they are, respectively, coupled to the left and right leads, and the two energy levels are also connected with coupling strength $t_{12}$ ($t_{21}=t_{12}^*$). In this system, we have the Rabi frequency between the two dots,
\begin{equation}
\Delta \omega = 2\sqrt{ \frac{\Delta \epsilon^2}{4} +|t_{12}|^2}
\ , \qquad  \Delta \epsilon =|\epsilon_1-\epsilon_2 | ,
\end{equation}
which is actually the difference between the two eigenvalues of the Hamiltonian of Eq. ({\ref{nr1}}).

Taking the band structure of the left and right leads into consideration, we assume that the leads have finite band-width in a Lorentzian form \cite{29} ${\bf \Gamma}_{\alpha}(\epsilon)=\frac{\Gamma_{\alpha} W^2}{\epsilon^2+W^2} $ where $\Gamma_{\alpha}$ is the linewidth amplitude of the left or right lead with $\Gamma_L=\Gamma_R=\Gamma/2$ and we further assume that both leads have the same bandwidth $W$. During the numerical calculation, the energies are measured in the unit of $\Gamma$ so that $1/\Gamma$ and $e\Gamma$ are the units of the time and current, respectively. In this paper, the bandwidth is chosen to be $W=10\Gamma$, the energy levels of the left and the right quantum dots are $\epsilon_1=6\Gamma$ and $\epsilon_2=4\Gamma$, respectively. At $t=0^-$ the system is disconnected. At $t=0^+$, the system is connected and the
Fermi level of the left lead is $\Delta_L=10\Gamma$ and the Fermi level of the right is zero.

For the double quantum dot system, the GF shall be written as
\begin{equation}
Z(\lambda ,t)=\det G \det \left(
\begin{array}{cc}
	g_1^{-1}-{\Sigma}_L   &  -t_{12}  \\
	-t_{21}   &   g_2^{-1}-\widetilde{\Sigma}_R
\end{array} \right),
\end{equation}
so that we are measuring electrons in the right lead. We also assume that the initial electron occupation of the energy level of the left quantum dot is zero and the initial occupation of the energy level of the right dot is one, then $g_1^{<}=0$ and $g_2^{<}(t_1,t_2)=i\exp[-i\epsilon_2(t_1-t_2)]$. The detailed description of the calculation of the GF which is actually a determinant in the time domain is presented in Appendix B.

In Fig.{\ref{fig4}}, we show the first-sixth cumulants of transferred charges which are counted from time $t_0=0$ to the time $t$ in the right lead of the system. The figure shows the cumulants as a function of time under different coupling strengths between the two dots with $t_{12}=1.5\Gamma, 3.0\Gamma$, and $6.0\Gamma$ at zero temperature, and we also show the influence of temperature on the cumulants at a temperature $k_B T=5\Gamma$ when the coupling strength $t_{12}=3.0\Gamma$, where $k_B$ is the Boltzmann constant. We can see from Fig.{\ref{fig4}} especially Figs.{\ref{fig4}}(e)-{\ref{fig4}}(f) that there are two kinds of oscillations in the cumulants: one is the overall oscillation, and the other one is the local oscillation with a specific period. Overall, there are more oscillations of the cumulants $\langle\langle n^j \rangle\rangle$ as one increases $j$, which shows the phenomenon of universal oscillations in FCS. The universal oscillations of the cumulants in the Coulomb blockade regime have been revealed experimentally by Flint \textit{et al.}.\cite{exp1} The local oscillation is caused by two serial quantum dots, since the electron in the quantum dots will oscillate between the two energy levels and the period of the local oscillations is $T_{osc}=2\pi/(2\Delta\omega)$.
The oscillation depends on the ratio of coupling strength between two dots and the coupling between the right dot and the right lead. If this ratio is small, the oscillation will not be so obvious, since it is easier for the electron in the right dot to tunnel to the right lead. This can be confirmed from Fig.{\ref{fig4}} that the oscillation of the cumulants of the system with a coupling strength $t_{12}=1.5\Gamma$ is weaker than the other two cases at zero temperature. However, if the coupling strength between the dots is strong enough, the first cumulant as in the case of
$t_{12}=6.0\Gamma$ in the figure may have negative values at short times, since the electron tends to oscillate between the dots and is unwilling to flow to the right lead.  This in turn creates a vacancy in the right dot and hence a larger possibility for the electron in the right lead to tunnel into the right dot giving rise to a negative current. It is found that the first and second cumulants, which are mean values and the variance, do not have too many local oscillations and are smooth at longer times.

Regarding the influence of the temperature, we compared the cumulants between zero temperature and $k_B T=5\Gamma$ when coupling strength $t_{12}=3.0\Gamma$. The temperature will reduce the probability that an electron transfer from the right quantum dot to the right lead and enhance the probability that an electron tunnel from the right lead to the right quantum dot. Both the overall oscillation and the local oscillation are smeared due to the temperature effect.

\begin{widetext}
\begin{figure*}
  \includegraphics[width=7.1in]{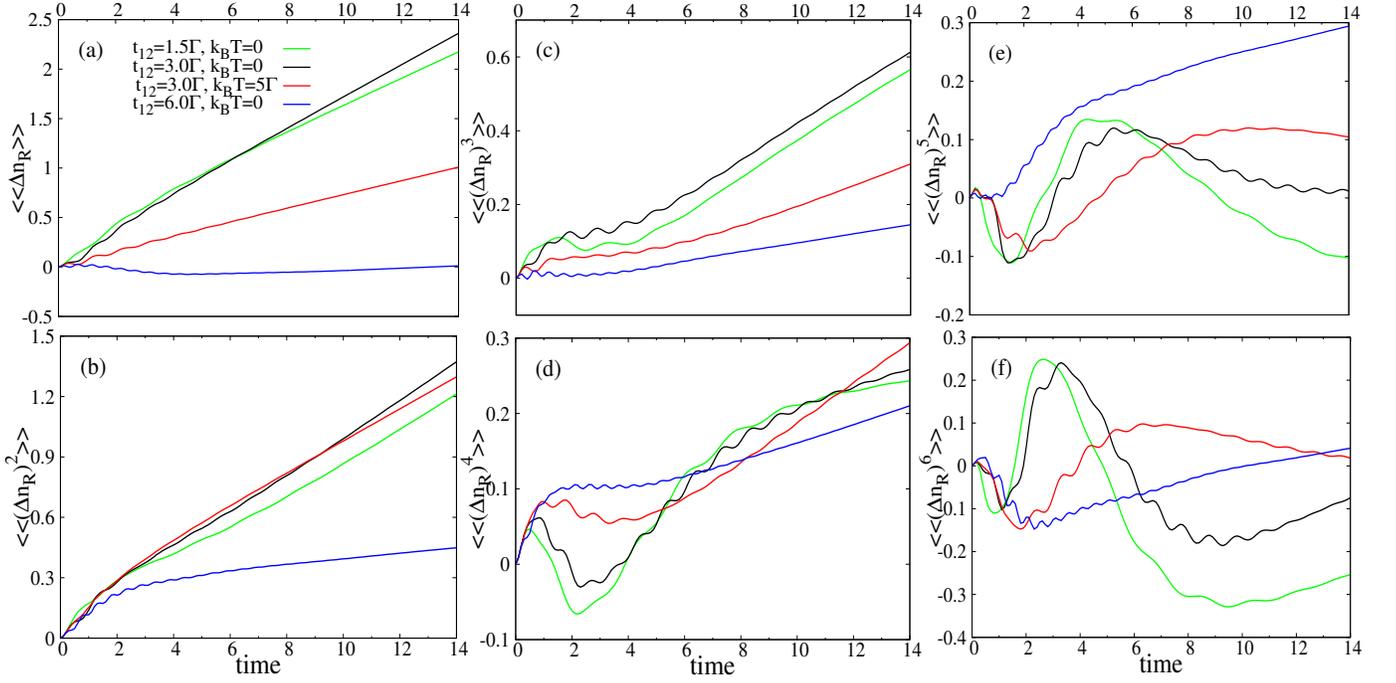} \\
  \caption{Cumulants ((a) 1st cumulant, (b) 2nd, (c) 3rd, (d) 4th, (e) 5th, (f) 6th) of transferred charges in the right lead of a system consisting of two single-level quantum dots connected to the left and right lead at time $t_0=0$. The numbers of the transferred charges are counted from time $t_0=0$ to $t$. The initial electron occupation of the energy level of the left quantum dot is zero and the initial occupation of the energy level of the right dot is one. $\hbar=e=\Gamma=1$, the energies are measured in the unit of $\Gamma$ and $1/\Gamma$  is the unit of time. The bandwidth is chosen to be $W=10\Gamma$, the energy levels of the left and the right quantum dot are $\epsilon_1=6\Gamma$ and $\epsilon_2=4\Gamma$, respectively. The figure shows the cumulants as a function of time at different coupling strengths between the two dots with $t_{12}=1.5\Gamma$, $3.0\Gamma$ and $6.0\Gamma$ at zero temperature, and we show the influence of temperature over the cumulants at a temperature $k_B T=5\Gamma$ when the coupling strength $t_{12}=3.0\Gamma$ as well, where $k_B$ is the Boltzmann constant.     }
  \label{fig4}
\end{figure*}
\end{widetext}

\begin{figure}
  \includegraphics[width=3.5in]{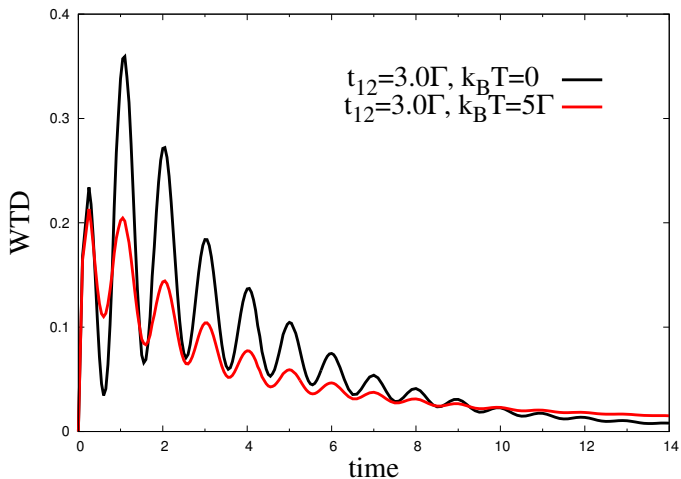}\\
  \caption{WTD ($W_1$) of the two quantum dot system. The initial electron occupation of the energy level of the left quantum dots is zero and the initial occupation of the energy level of the right dot is one. The bandwidth is chosen to be $W=10\Gamma$, the energy levels of the left and the right quantum dot are $\epsilon_1=6\Gamma$ and $\epsilon_2=4\Gamma$, respectively. We compare WTD at $k_B T=0$ and $5\Gamma$ with the coupling strength $t_{12}=3.0\Gamma$.    }
  \label{fig5}
\end{figure}

In Fig. {\ref{fig5}}, we calculated the WTD ($W_1$) in the right lead in the transient regime, which is the probability distribution that the first electron transfer to the right lead at different times after we turn on the interaction between the leads and the quantum dots at $t=0$. The WTD of the system with parameters $\Delta \epsilon=2\Gamma ,\ t_{12}=3.0\Gamma$ at zero temperature and $\ k_B T=5\Gamma$ are presented. Except from the first peak of each curve, we can see from Fig.{\ref{fig5}} that WTD exhibit an oscillation with a period $T_{osc}=2\pi/(2\Delta\omega)$ again due to Rabi oscillation. The temperature does not influence the oscillation period but it smears the oscillation amplitude since temperature only influences the electronic distribution in two leads.
	

\section{Conclusion}

Using the technique of path integral and Keldysh nonequilibrium Green's function, we express the GF in a compact form in terms of Green's function and self-energy in the time domain. This formalism is suitable for studying FCS in the transient regime. For the dc steady state regime, two measurements are needed to collect to investigate the finite-time FCS. As we have shown in this paper, the first measurement actually perturbs the system and hence FCS after the measurement does not reflect information of real system. Therefore, this formalism can not be used to study finite-time FCS for the dc steady state. We have generalized the formalism to the magnetic tunnel junction to study FCS of spin-polarized charge current, spin current, and spin-transfer torque. Moreover, we have calculated GF for the quantum point contact system in the transient regime. We have applied our theory to study FCS of a double quantum dot system. Both global and local oscillations are revealed. We attribute the global oscillation to the universal oscillation as observed experimentally in the Coulomb blockade regime. The local oscillation can be understood from the Rabi oscillation.

Future work may involve transient FCS of charge transport in quantum point contact systems such as conductor-superconductor (N-S) and superconductor-superconductor (S-S) systems. In addition, the transient FCS of spin transport in a mesoscopic system with spin-orbit interaction is also worth studying. 

Finally, we note that the theoretical framework presented here can not be applied in the presence of strong electron-electron interactions. Although exact result cannot be obtained, we think that the perturbative approach can be used in dealing with the interactions. This is an interesting research topic which we will pursue in the near future.

\begin{acknowledgements}
We thank J.-S. Wang for useful discussions. This work was financially supported by the RGC (Grant No. HKU 705212P) and the UGC (Contract No. AoE/P-04/08) of the Government of the HKSAR, the National Natural Science Foundation of China (Grant No. 11374246).
\end{acknowledgements}

\section*{APPENDIX A: FERMIONIC COHERENT STATES}
The fermionic coherent states are defined in terms of the linear superposition of the vacuum state $|0\rangle$ and occupied state $|1\rangle$ parametrized by two unrelated complex numbers $\phi$ and $\overline{\phi}$ which are called Grassmann variables,\cite{37}
\begin{equation} \label{eqA1}
|\phi\rangle \equiv |0\rangle -\phi|1\rangle
=(1- \phi c^{\dag})|0\rangle   \tag{A1}
\end{equation}
\begin{equation} \label{eqA2}
\langle\phi| \equiv \langle 0|-\langle 1|\overline{\phi}
=\langle 0|(1-c\overline{\phi}) .     \tag{A2}
\end{equation}
The coherent states are the eigenstates of the annihilation operator:
\begin{equation} \label{eqA3}
c|\phi\rangle = \phi |\phi\rangle .  \tag{A3}
\end{equation}
Similarly,
\begin{equation}  \label{eqA4}
\langle\phi|c^{\dag} =\langle\phi| \overline{\phi} . \tag{A4}
\end{equation}
The Grassman variables satisfy the following equations:
\begin{equation}  \label{eqA5}
(\phi)^2=(\overline{\phi})^2=0 ,  \qquad \{\phi,\overline{\phi} \}_+=0 . \tag{A5}
\end{equation}

From Eq.~(\ref{eqA5}) we know that any function of the Grassmann algebra is at most of the second-order
\begin{equation}    \label{eqA6}
f(\phi,\overline{\phi})=A+B\phi +C\overline{\phi}+D\phi\overline{\phi} .  \tag{A6}
\end{equation}

Integrations of the Grassmann variables are defined as
\begin{equation}  \label{eqA7}
\int d\phi \ 1=\int d\overline{\phi} \ 1=0, \qquad
\int d\phi \ \phi =\int d\overline{\phi} \ \overline{\phi}=1.  \tag{A7}
\end{equation}
Differentials of the Grassmann variables are defined as
\begin{equation}  \label{eqA8}
\frac{\partial}{\partial\phi} f(\phi,\overline{\phi})=B+D\overline{\phi},  \qquad
\frac{\partial}{\partial\overline{\phi}} f(\phi,\overline{\phi})=C-D\phi.   \tag{A8}
\end{equation}
This implies that
\begin{equation}  \label{eqA9}
\frac{\partial}{\partial\phi}\frac{\partial}{\partial\overline{\phi}}
=-\frac{\partial}{\partial\overline{\phi}}\frac{\partial}{\partial\phi}  . \tag{A9}
\end{equation}
Performing the integral of $f(\phi,\overline{\phi})$ with respect to $\phi$ or $\overline{\phi}$ and comparing with Eq.~(\ref{eqA8}), we obtain the operator identities
\begin{equation}  \label{eqA10}
\int d\phi=\frac{\partial}{\partial\phi},  \qquad
\int d\overline{\phi}=\frac{\partial}{\partial\overline{\phi}} .  \tag{A10}
\end{equation}

Using Eqs.~(\ref{eqA7}) and~(\ref{eqA10}), we obtain the functional Gaussian integral for the Grassmann variables for any invertible complex $N\times N$ matrix $M$:
\begin{align}  \label{eqA11}
&\int \mathcal{D}(\overline{\phi}\phi)\exp\left[- \sum_{i,j}\overline{\phi}_i M_{ij} \phi_j+ \overline{\kappa}_i\phi_i+ \overline{\phi}_i\kappa_i \right] \notag \\
=& \det M \exp\left[ \sum_{ij}
\overline{\kappa}_i(M^{-1})_{i,j}\kappa_j \right] ,
 \tag{A11}
\end{align}
where $\mathcal{D}(\overline{\phi}\phi)=\prod_{i=1}^{N}d\overline{\phi}_id\phi_i$. If we set $\overline{\kappa}_i=\kappa_i=0$, we arrive at $\int \mathcal{D}(\overline{\phi}\phi)\exp\left[- \sum_{i,j}\overline{\phi}_i M_{ij} \phi_j \right] =\det M $.

Using Eqs.~(\ref{eqA1}),~(\ref{eqA2}), and~(\ref{eqA5}), we find the overlap between any two coherent states as
\begin{equation}  \label{eqA12}
\langle \phi| \phi^{\prime}\rangle =1+\overline{\phi}\phi^{\prime} =\exp\{\overline{\phi}\phi^{\prime}\}  \tag{A12}  .
\end{equation}

From Eqs.~(\ref{eqA3}),~(\ref{eqA4}), and~(\ref{eqA6}), the matrix elements of a {\it normally ordered} operator, such as the Hamiltonian, take the form
\begin{equation}  \label{eqA13}
\langle \phi| H(c^{\dag},c)|\phi^{\prime}\rangle =H(\overline{\phi},\phi^{\prime}) \langle \phi| \phi^{\prime}\rangle =H(\overline{\phi},\phi^{\prime}) \exp\{\overline{\phi}\phi^{\prime}\}  . \tag{A13}
\end{equation}
Similarly,\cite{foot3}
\begin{equation}  \label{eqA14}
\langle \phi| e^{\kappa c^{\dag}c}|\phi^{\prime}\rangle =\exp\{\overline{\phi}\phi^{\prime} e^{\kappa}\} .     \tag{A14}
\end{equation}

The differential elements $d\phi$ and $d\overline{\phi}$ anticommute with each other. Using Eqs.~(\ref{eqA1}),~(\ref{eqA2}),~(\ref{eqA7}), and~(\ref{eqA12}), it is straightforward for us to get the over-completeness of the fermion coherent state
\begin{equation}  \label{eqA15}
1=\int d\overline{\phi} d\phi \exp(-\overline{\phi}\phi)|\phi \rangle \langle \phi|
 =\int d\phi d\overline{\phi} \exp(\overline{\phi}\phi)|\phi \rangle \langle \phi| .
 \tag{A15}
\end{equation}
The trace of an operator, $\hat{A}$, is calculated as:
\begin{equation}  \label{eqA16}
\mathbf{Tr}\hat{A}=\int\int d\overline{\phi} d\phi e^{-\overline{\phi}\phi}\langle \phi |\hat{A}|-\phi\rangle  .
 \tag{A16}
\end{equation}

\section*{APPENDIX B: Numerical details}
Here we present detailed description on how to calculate the generating function which is a functional determinant described by Green's function and self-energy in the transient regime. Since the functional determinant is expressed in the time domain, we should make a discretization of the time indices. The determinant can be calculated through Eq.(\ref{eq62}), and we should keep in mind that both the Green's functions and self-energies have different Keldysh components. The Green's function can be obtained through the Dyson equation, which is Eq.(\ref{eq58}) on the matrix level. For the retarded Green's function, we should first discretize $G^r, \ g^r$, and $\Sigma^r$ which have two time indices with a time slice $\Delta t$, and by the rule of matrix multiplication, we have
\begin{equation*}
\underline{G}^r=\underline{g}^r +\underline{g}^r \underline{\Sigma}^r \underline{G}^r \Delta t^2  ,
\end{equation*}
where we have used the underlined Green's function and self-energy to denote the Green's function and self-energy in the matrix form. Given the self-energy and the Green's function $g$ of the isolated central system, we can calculate the Green's function of the system using
\begin{equation*}
\underline{G}^r=(I-\underline{g}^r\underline{\Sigma}^r \Delta t^2)^{-1} \underline{g}^r  ,
\end{equation*}
where $I$ is the identity matrix. From Eq.(\ref{eq62}), we obtain $\underline{G}^<$ which allows us to calculate the generating function $Z(\lambda ,t)$. However, this method is time consuming, since at every time step, we should do a matrix inversion to get $\underline{G}^r$.

Following, we introduce a method to make the calculation much more efficient.  First we calculate the isolated Green's function of the central system and self-energy with different Keldysh components in the time domain.\cite{Yuzhu}	
For the quantum dot with single energy level $\epsilon_0$, $g^r(\tau_1,\tau_2)=-i\theta(\tau_1-\tau_2)\exp[-i\epsilon_0(\tau_1-\tau_2)]$, where $\theta(\tau_1-\tau_2)$ is the Heaviside step function and $g^a$ is the Hermitian conjugate of $g^r$. $g^{<}(\tau_1,\tau_2)$ is zero if the initial occupation of the energy level is empty while if the energy level is initially occupied with one electron $g^{<}(\tau_1,\tau_2)=i\exp[-i\epsilon_0(\tau_1-\tau_2)]$. Then $g^k(\tau_1,\tau_2)$ is found through the relation $g^k=2g^<+g^r-g^a$.

	 The equilibrium self-energies are chosen to be energy dependent with a finite band width $W$,
\begin{equation}  \label{eqB1}
\bar{\Sigma}_{\alpha}^r(\omega)=\frac{\Gamma_{\alpha} W}{2(\omega+iW)} , \tag{B1}
\end{equation}
so that the linewidth function is the following Lorentzian form:
\begin{equation}  \label{eqB2}
{\bf \Gamma}_{\alpha}(\epsilon)=\frac{\Gamma_{\alpha} W^2}{\epsilon^2+W^2}  \tag{B2}
\end{equation}
where $\Gamma_{\alpha}$ is the linewidth amplitude.
The self-energy in the time domain is defined as
\begin{equation}   \label{eqB3}
\Sigma_{\beta}^{r,<}(\tau_1,\tau_2)=\int \frac{d\omega}{2\pi} e^{-i\omega(\tau_1-\tau_2)}\bar{\Sigma}_{\beta}^{r,<}(\omega) e^{ -i\int_{\tau_2}^{\tau_1}\Delta_{\beta}(t) dt }   \tag{B3}
\end{equation}
where $\bar{\Sigma}_{\beta}^{r,<}$ is the equilibrium self-energy in the energy domain and $\Delta_\beta$ is the external bias voltage in the lead $\beta$. Using Eq.(\ref{eq3}), we find the retarded self-energy of the left lead:
\begin{equation}   \label{eqB4}
\Sigma_L^r(\tau_1,\tau_2)=-\frac{i}{4}
\theta(\tau_1-\tau_2)\Gamma W e^{-(i\Delta_L+W)(\tau_1-\tau_2)}   \tag{B4}
\end{equation}
where we have assumed $\Gamma_L=\Gamma_R=\Gamma/2$.
For the lesser self-energy
\begin{equation}    \label{eqB5}
\Sigma_L^<(\tau_1,\tau_2)=i\int\frac{d\omega}{2\pi} e^{-i\omega(\tau_1-\tau_2)}
e^{-i\Delta_L(\tau_1 -\tau_2)}f(\omega){\bf \Gamma}_L(\omega)   \tag{B5}
\end{equation}
with $f(\omega)=1/\left[e^{\beta(\omega-E_F)}+1\right]$ and $E_F=0$.


At zero temperature, note the following:
\begin{align}    \label{eqB6}
\Sigma_L^<(\tau_1,\tau_2)
&=i e^{-i\Delta_L(\tau_1 -\tau_2)} \int_{-\infty}^{0}\frac{d\omega}{2\pi} e^{-i\omega(\tau_1-\tau_2)}\frac{\Gamma_L W^2}{\omega^2+W^2}   \tag{B6}
\end{align}
 \\
1. If $\tau_1=\tau_2$,
\begin{equation}    \label{eqB7}
\Sigma_L^<(\tau_1,\tau_2)=\frac{i}{8}\Gamma W    \tag{B7}
\end{equation}
2. If $\tau_1>\tau_2$, let $\tau=\tau_1-\tau_2$,
\begin{align}    \label{eqB8}
\Sigma_L^<(\tau_1,\tau_2) &=\frac{i}{8}\Gamma W
\left\{\frac{i}{\pi} e^{(W-i\Delta_L)\tau} E1(W\tau)  \right. \notag \\
&\left.  +e^{-(W+i\Delta_L)\tau} \left[2-\frac{i}{\pi}E1(-W\tau) \right]\right\}
 \tag{B8}
\end{align}
where $E1(x)=\int_x^{\infty}\frac{e^{-t}}{t}dt$.

At non-zero temperature,

1. if $\tau_1=\tau_2$, the integral is actually Hilbert transformation of the Fermi distribution function.\cite{36}
\begin{equation}   \label{eqB9}
\Sigma_L^<(\tau_1,\tau_2)=\frac{i\Gamma W}{8}   \tag{B9}
\end{equation}
2. if $\tau_1>\tau_2$, it has poles $\frac{-i(2n+1)\pi}{\beta}$ and $-iW$, where $n=0,1,2,3...$, we have
\begin{align}  \label{eqB10}
&\Sigma_L^<(\tau_1,\tau_2)=
\frac{i\Gamma_L W\exp[-(W+i\Delta_L)(\tau_1-\tau_2)]}{2\exp(-i\beta W)+2}
-\frac{1}{\beta} \times    \notag \\
&\sum_{n=0}^{+\infty}\exp\left\{-[\frac{(2n+1)\pi}{\beta}+i\Delta_L](\tau_1-\tau_2) \right\}\frac{\Gamma_L W^2}{W^2-[\frac{(2n+1)\pi}{\beta}]^2} .   \tag{B10}
\end{align}
Using the relation $\Sigma_L^<(\tau_1,\tau_2)\big|_{\tau_1<\tau_2}=-[\Sigma_L^<(\tau_1,\tau_2)\big|_{\tau_1>\tau_2}]^*$, we obtain the full expression of $\Sigma_L^<(\tau_1,\tau_2)$. The expression of $\Sigma_R^<(\tau_1,\tau_2)$ can be obtained similarly. Finally, using the relation $\Sigma^k=2\Sigma^<+\Sigma^r-\Sigma^a$, we could know $\Sigma^k(\tau_1, \tau_2)$.

	We know that a contour ordered matrix $A$ could be written in the upper triangular form $\left( \begin{array}{cc}
A^r  & A^k  \\  0 &  A^a	
\end{array}	 \right)$ in the Keldysh space after Keldysh rotation.  Since $G$, which does not contain the counting parameter, possesses the upper triangular form in Keldysh space, and its retarded and advanced components are lower triangular and upper triangular matrices, respectively, in the time domain, we can just simplify it to a diagonal matrix. So, we can just directly calculate GF by calculating the determinant of the matrix $\delta(g^{-1}-\widetilde{\Sigma}_L-\Sigma_R)$ which is a block toeplitz matrix where $\delta$ is the diagonal matrix to satisfy the normalization condition $Z(\lambda=0,t)=1$.


\begin{thebibliography}{99}
\bibitem{foot1}
The last equality in this equation can be shown by splitting $\exp(i\lambda n_0)$ into two $\exp(i\lambda n_0/2)$ and inserting them at the first and second $P_0$, respectively.
\bibitem{foot2}
Because of the negative sign in front of the spin-down current in Eq.(\ref{mtj16}), we need to put a minus sign in the counting field to reflect this fact. As a result, we have Eq.(\ref{mtj18}). The extra factor of $1/2$ in the definition of ${\bar \Lambda}$ is due to the $s=1/2$ for spin.
\bibitem{foot18}
Eq.(15) of Ref.\onlinecite{YunjinYu}) can be obtained from Eq.(\ref{mtj19}) after the Bogoliubov transformation.
\end{thebibliography}
\end{document}